\documentclass[letterpaper,12pt]{iopart}
\expandafter\let\csname equation*\endcsname=\relax
\expandafter\let\csname endequation*\endcsname=\relax
\usepackage{amsmath, color, amssymb, graphicx, stmaryrd, wasysym}
\definecolor{linkcolor}{rgb}{0,0,0.6} 
\usepackage[colorlinks=true,
	pdfstartview = FitV,
	linkcolor    = linkcolor,
	citecolor    = linkcolor,
	urlcolor     = linkcolor,
	hyperindex   = true,
	hyperfigures = false]{hyperref}

\usepackage{amsmath,amssymb,graphicx,cases}
\usepackage{epsfig}
\usepackage{textcomp}
\usepackage{epstopdf}
\usepackage{float}
\usepackage{braket}
\usepackage[normalem]{ulem}
\usepackage{enumerate}
\usepackage{enumitem}
\usepackage{ae,aecompl}
\usepackage{lmodern}
\usepackage{physics}

\usepackage{xcolor,cancel}
\definecolor{mygreen}{rgb}{0.0,0.55,0.4}

\newcommand{\stkout}[1]{\ifmmode\text{\sout{\ensuremath{#1}}}\else\sout{#1}\fi}

\begin{document}

\title[Irreversibility in scalar active turbulence: The role of topological defects]{Irreversibility in scalar active turbulence: The role of topological defects}

\author{Byjesh N. Radhakrishnan$^{*}$, Francesco Serafin$^{*\dagger}$,\\ Thomas L. Schmidt, and \'Etienne Fodor}
\address{Department of Physics and Materials Science, University of Luxembourg, L-1511 Luxembourg, Luxembourg}

\begin{abstract}
In many active systems, swimmers collectively stir the surrounding fluid to stabilize some self-sustained vortices. The resulting nonequilibrium state is often referred to as active turbulence. Although active turbulence clearly operates far from equilibrium, it can be challenging to pinpoint which emergent features primarily control the deviation from an equilibrium reversible dynamics. Here, we reveal that dynamical irreversibility essentially stems from singularities in the active stress. Specifically, considering the coupled dynamics of the swimmer density and the stream function, we demonstrate that the symmetries of vortical flows around defects determine the overall irreversibility. Our detailed analysis leads to identifying specific configurations of defect pairs as the dominant contribution to irreversibility.
\end{abstract}
\vspace*{\fill}
\scalebox{.8}{$^*$These authors contributed equally to this work.}\\
\scalebox{.8}{$^\dagger$E-mail: francesco.serafin@uni.lu}
\maketitle


\section{Introduction}

In active systems, each unit extracts energy from and transfers work to its surroundings~\cite{Marchetti2013, FODOR2018}. The violation of time-reversal symmetry (TRS) at the microscale yields nonequilibrium collective states; for instance, collective directed motion~\cite{Chate2020}, motility-induced phase separation (MIPS)~\cite{Cates_2015}, and spatiotemporal chaotic flows referred to as {\it active turbulence}~\cite{Alert_2022}. From a top-down perspective, continuum theories based on symmetry arguments, which are inspired by equilibrium theories~\cite{Hohenberg_1977}, capture nonequilibrium states by considering additional terms breaking TRS~\cite{Wittkowski_2014, nardini_2017, Markovich2021}. In the presence of a momentum-conserving fluid, wet theories describe the coupling between the emergence of order, for instance nematics~\cite{Doostmohammadi2018}, and the advection of large-scale flows.

Active turbulence has attracted increasing attention due to similarities with passive turbulence, such as power-law scaling of the flow velocity fluctuations~\cite{Alert_2021}. Yet, recent works have revealed that active turbulence does not require any inertia and does not feature any energy cascade~\cite{Alert2020}, in contrast with its passive counterpart. Remarkably, various active systems can feature active turbulence: bacterial suspensions~\cite{Dombrowski_2004, Henricus_2012, Dunkel_2013, Patteson_2018,liu_2021}, swarming sperms~\cite{Creppy_2015}, motility assays~\cite{Sanchez_2012, Guillamat_2017, Lemma_2019, Tan_2019}, cell monolayers~\cite{Doostmohammadi_2015, Blanch_2018, Lin_2021}~and suspensions of self-propelled colloids~\cite{Doostmohammadi_2015, Lin_2021, Blanch_2018}. A distinct attribute of active turbulence is the sustained creation and annihilation of topological defects: these singularities correspond to disclinations in the nematic texture which are stirred by vortical flow~\cite{Giomi_2013, thampi_2013, Giomi_2015, Doostmohammadi_2015, Doostmohammadi2018}. Eﬀective interactions between defects can be described as being mediated by the interplay between elastic distortions of the nematic texture and active ﬂows \cite{Giomi_2013, Shankar_2018, Shankar_2019}.

Quantifying TRS breakdown is a way to measure the deviation from equilibrium~\cite{Lebowitz_1999, Maes_1999, Jorge_1998}. In fact, TRS is the essential symmetry of passive dynamics~\cite{Onsager1931}, from which all equilibrium properties follow. In active matter, identifying the regime of weak TRS is a route towards a perturbative analysis of nonequilibrum features close to equilibrium~\cite{Fodor2016, Borthne_2020, Martin2021, Loos2023, Bertrand2023, Agranov2024, Agranov2025, Proesmans2025}. For thermodynamically-consistent active models~\cite{Gaspard2019, Markovich2021, Aslyamov2023, Bebon2024, Agranov2024, chatzittofi_entropy_2024, Cocconi2025}, quantifying TRS breakdown provides access to the amount of energy dissipated into the thermostat, which coincides with the entropy production in steady state~\cite{Seifert_2012, peliti2021stochastic}. In most active models, where the dynamics of underlying (chemical) degrees of freedom is discarded, although measuring irreversibility underestimates dissipation, it leads to identifying near-equilibrium regimes. To this end, one can define an {\it informatic entropy production rate} (IEPR) by comparing the probability of forward and time-reversed trajectories~\cite{Jack2022}.

To elucidate the relation between TRS breakdown and nonequilibrium patterns, it is convenient to introduce a local decomposition of IEPR~\cite{Jack2022}. For MIPS~\cite{Cates_2015}, this approach has revealed the essential role of nonequilibrium forces in shaping the interface between dilute and dense regions~\cite{nardini_2017, Martin2021}, in line with thermodynamically consistent models~\cite{Markovich2021, Bebon2024}. In a similar fashion, the decomposition of IEPR can be examined in a broad class of active hydrodynamics theories. For instance, recent works have shown how fluid flows can destabilize active phase separation, yielding a regime akin to active turbulence~\cite{cates2015, Singh2019, Das2020, Bhattacharjee2022, Padhan2024}. In this context, it remains to explore whether the local IEPR can be related to any striking features of the self-sustained flows.

In this paper, we examine the interplay between IEPR and topological defects {(TDs)} in a continuum model of active turbulence. Inspired by previous works~\cite{cates2015, Singh2019, Das2020, Bhattacharjee2022, Padhan2024}, our model describes the density dynamics of swimmers immersed in a momentum-conserving fluid. Recently,  In analogy with the turbulence of active nematics~\cite{Alert2020}, we show that vortical flows can spontaneously form despite the absence of inertia. By analytically deriving the IEPR, we demonstrate that the dominant contribution to irreversibility stems from vortices. We also show that the IEPR is a lower bound of the total dissipation derived from linear irreversible thermodynamics (LIT). Finally, we propose a mapping between the shear stress, which stirs the fluid, and an effective nematic field. Such a mapping leads to delineating singularities in the density profile which shape the surrounding vortices. In all, our results reveal how the symmetries of vertical flows, set by topological defects, determine the spatial distribution of IEPR.

The paper is structured as follows. In Sec.~\ref{sec:model}, we start by introducing our continuum theory, and reviewing the associated phenomenology: laminar, vortex, and turbulent states. We then show that the zeros of density gradients correspond to nematic-like singularities, and numerically evaluate the corresponding defect statistics. In Sec.~\ref{sec:epr}, we explicitly relate the IEPR to the vorticity, and provide a numerical quantification in the noiseless regime. We also find a generic {relation} between the IEPR and the total dissipation. We find that in steady state the IEPR is a lower bound to the total EPR. Finally, by computing the vorticity around defects, we examine in detail how the spatial distribution of IEPR is determined by the symmetries of  defects.


\section{Turbulence in Active Model H}\label{sec:model}

In this section, we present the dynamics of Active Model H (AMH) \cite{cates2015} in the absence of inertia. This dynamics couples two fields: (i)~a conserved scalar field, representing the local density of active swimmers, and (ii)~a non-conserved scalar field corresponding to the stream function of the surrounding fluid flow. We characterize the emergence of three main states: laminar flow, vortex array, and turbulence. We discuss the spontaneous formation of topological defects by analyzing the profile of density gradients.

\subsection{Coupled dynamics of density and stream function}
\label{subsec:active_model_h}

Inspired by previous works~\cite{cates2015, Singh2019}, we start by considering the dynamics of the local density of active swimmers $\phi(\vb*{r},t)$ coupled to  the velocity $\vb*{v}(\vb*{r},t)$ of a momentum-conserving fluid
\begin{equation}\label{eq:phi_dynamics}
	\dot{\phi}+\vb*{v} \cdot \nabla \phi =-\nabla \cdot \vb*{J} ,
	\quad
	\vb*{J} =-M\nabla \frac{\delta \mathcal{F}}{\delta \phi} + \sqrt{2MT} \vb*{\Lambda}_{\phi} ,
\end{equation}
where $M$ is the mobility, and $T$ the temperature  (we use units in which $k_B=1$). We set $M=1$ in the following. The noise $\vb*{\Lambda}$ is Gaussian with zero mean and correlations given by
\begin{equation}
	\langle \Lambda_{\phi,\alpha}(\vb*{r},t) \Lambda_{\phi,\beta}(\vb*{r}',t') \rangle = \delta_{\alpha\beta} \delta(\vb*{r}-\vb*{r}') \delta (t-t') .
\end{equation}
We take for the free-energy functional $\cal F$ a form analogous to a Landau-Ginzburg expansion:
\begin{equation}\label{eq:free}
	\mathcal{F}[\phi]=\int \left [\frac{\kappa}{2}(\nabla \phi)^{2} + g(\phi)\right ]\mathrm{d}\vb*{r} ,
	\quad
	g(\phi)=\frac{a}{2}\phi^{2}+\frac{b}{4}\phi ^{4} ,
\end{equation}
where $(\kappa,a,b)$ are parameters. A phase separation emerges for $a < 0$ and $b,\kappa>0$ ~\cite{Wittkowski_2014, Bray_1994, cates_2013}. The density dynamics in Eq.~\eqref{eq:phi_dynamics} must be supplemented with the Navier-Stokes equation for the velocity dynamics:
\begin{equation}\label{eq:Naverthe Stokes}
	\rho(\partial_tv_\alpha+v_\beta\partial_\beta v_\alpha) = \eta \nabla^2 v_\alpha -\partial_\alpha p +\partial_{\beta}\left (\Sigma _{\alpha \beta}^{P}+\Sigma_{\alpha \beta}^{A}+\sqrt{2\eta T}\Gamma_{\alpha \beta}\right ) ,
\end{equation}
where $\rho$ is the fluid's density, $p(\vb*{r},t)$ the pressure field, and $\eta$ the fluid's viscosity. Summation over repeated indices is assumed throughout the paper. The fluid's density obeys the continuity equation $\partial_t\rho + \partial_\beta(\rho v_\beta)=0$. Here, $\Gamma_{\alpha\beta}$ is a Gaussian noise field with zero mean and correlations given by~\cite{Zwanzig}
\begin{equation}
	\langle{\Gamma_{\alpha\beta}(\vb*{r},t)\Gamma_{\mu\nu}(\vb*{r}',t')}\rangle = (\delta_{\alpha\mu}\delta_{\beta\nu}+\delta_{\alpha\nu}\delta_{\beta\mu})\delta(\vb*{r}-\vb*{r}')\delta(t-t') .
\end{equation}
The passive and active contributions to the deviatoric stress, respectively $\Sigma_{\alpha \beta}^{P}$ and $\Sigma_{\alpha \beta}^{A}$, read
\begin{equation}\label{eq:stress_tensor}
	\Sigma_{\alpha \beta}^{P} = -\kappa\left [ (\partial _{\alpha}\phi)(\partial _{\beta}\phi) - \frac{1}{2}|\nabla \phi|^{2} \delta_{\alpha \beta}\right ] ,
	\quad
	\Sigma_{\alpha \beta}^{A} = -\zeta\left[ (\partial _{\alpha}\phi)(\partial _{\beta}\phi) - \frac{1}{2}|\nabla \phi|^{2} \delta_{\alpha \beta}\right ] ,
\end{equation}
where $\zeta$ is the activity parameter~\cite{cates2015}. In passive systems, the expression of $\Sigma_{\alpha \beta}^{P}$ can be derived from the free-energy functional $\mathcal{F}[\phi]$~\cite{Chaikin_Lubensky_1995}. To the lowest order in powers of $\phi$ and its gradient, the term $\partial _{\alpha}\phi\partial _{\beta}\phi-\frac{1}{2}|\nabla \phi|^{2} \delta_{\alpha \beta}$ is the only traceless symmetric tensor that can be constructed, so that we assume that $\Sigma_{\alpha \beta}^{A}$ must be proportional to it. The sign of $\zeta$ distinguishes the cases of contractile ($\zeta< 0$) and extensile ($\zeta >0$) swimmers. The active stress is the only source of activity in the dynamics. Beyond fluid-like systems, recently it was proposed that the active stress of AMH could play a role in active tissues by mediating the coupling between a scalar concentration of proteins and the mechanical stresses in the tissue~\cite{MerkelNJP2023,MerkelArXiv2024,Merkel_NatPh_2025}.

In the following, we assume that the system is overdamped, and neglect the inertial term in Eq.~\eqref{eq:Naverthe Stokes}. The fluid's density decouples from the other fields. We further assume that the fluid is incompressible and we neglect inertial forces, yielding the Stokes equation:
\begin{equation}\label{eq:stokes}
	\eta \nabla^{2}v_{\alpha}=\partial _{\alpha}p-\partial_{\beta}\left (\Sigma _{\alpha \beta}^{P} + \Sigma_{\alpha \beta}^{A}+\sqrt{2\eta T}\Gamma_{\alpha \beta}\right ) ,
	\quad 
	\nabla \cdot \vb*{v}=0 \, ,
\end{equation}
{where
\begin{equation}\label{eq:fa}
    f^A_\alpha = \partial_\beta \Sigma_{\alpha \beta}^{A}
\end{equation}
is the active force and $f_\alpha = \partial_\beta (\Sigma^P_{\alpha \beta}+\Sigma^A_{\alpha \beta})$ is the total force.} To eliminate the pressure contribution in Eq.~\eqref{eq:stokes}, we define the stream function $\psi(\vb*{r},t)$ as 
\begin{equation}\label{eq:streamF}
	v_{x} = \partial _{y}\psi ,
	\quad
	v_{y} = -\partial _{x}\psi ,
\end{equation}
or, $v_\alpha=\epsilon_{\alpha\beta}\partial_\beta\psi$ (where $\epsilon_{\alpha\beta}=-\epsilon_{\beta\alpha}$, $\epsilon_{xy}=+1$ is the Levi-Civita symbol in 2 dimensions). By taking the curl of Eq.~\eqref{eq:stokes} and using Eq.~\eqref{eq:streamF}, the Stokes equation reduces to
\begin{equation}\label{eq:stream_dynamics}
	\eta \nabla ^{4} \psi =\left (\kappa+\zeta \right )[(\partial_x \phi)\nabla ^2 (\partial_y \phi)-(\partial_y \phi)\nabla ^2 (\partial_x \phi)]-\sqrt{2\eta T}\nabla ^2 \Lambda_\psi ,
\end{equation}
where $\Lambda_{\psi}$ is a scalar Gaussian noise with zero mean and correlations given by~[\ref{ap:stream_derivation}]
\begin{equation}
	\langle{\Lambda_\psi(\vb*{r},t)\Lambda_\psi(\vb*{r}',t')}\rangle = \delta(\vb*{r}-\vb*{r}'){\delta}(t-t') .
\end{equation}
In Eq.~\eqref{eq:stream_dynamics}, the vorticity $\omega=-\nabla^2\psi$ is sourced by a combination of passive and active stresses. The latter stir the system out of equilibrium and can generate spatiotemporal chaotic flow~\cite{cates2015}. In what follows, we analyze the field dynamics of AMH as governed by Eqs.~\eqref{eq:phi_dynamics} and \eqref{eq:stream_dynamics}.


\subsection{Laminar, vortex and turbulent states}\label{sec:steady_states}

We numerically simulate AMH [Eqs.~\ref{eq:phi_dynamics} and \ref{eq:stream_dynamics}] with the integration scheme detailed in \ref{ap:numerical_method}. The parameters are fixed to $-a=b=0.1$, $\kappa=0.1$, and $\eta=1.67$, and we vary the activity parameter $\zeta<0$, namely, considering contractile activity. The extensile case ($\zeta>0$) does not have turbulent phases, so it is not relevant for our study which focuses on the turbulent, defect-rich phase.

\begin{figure}
	\centering
	\includegraphics[width=1.\linewidth]{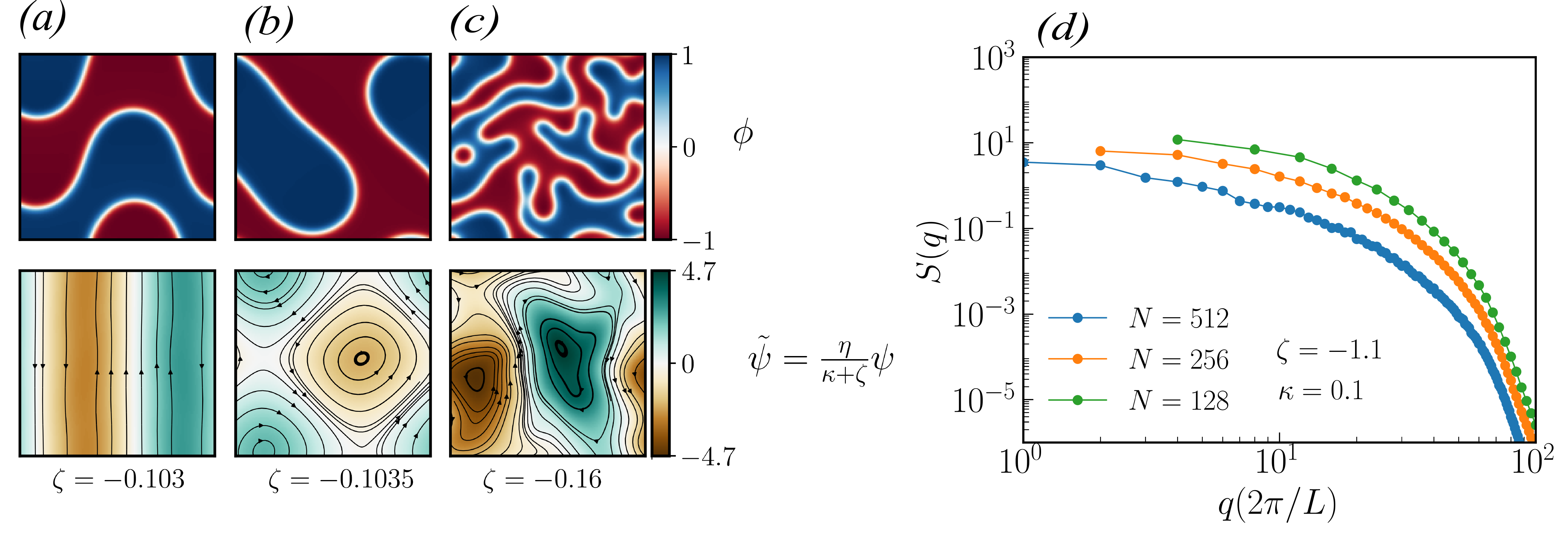}
	\caption{For Active Model H [Eqs.~\eqref{eq:phi_dynamics} and \eqref{eq:stream_dynamics}], the swimmer density $\phi$ (top) and the scaled stream function $\tilde{\psi} = \frac{\eta}{\zeta+\kappa} \psi $ (bottom) feature three possible steady states for different levels of activity $\zeta$: (a)~laminar flow, (b)~vortex, and (c)~spatio-temporal chaos (referred to as the turbulent state). The black lines with arrows show the velocity field lines. Parameters $-a=b=0.1$, $\kappa=0.1$, and $\eta=1.67$.
	(d)~Velocity spectrum $S(q)$ [Eq.~\eqref{eq:S}] as a function of wavenumber $q$ (scaled by the largest system size) for $\zeta =-1.1$.
	}
	\label{fig:steady_states}
\end{figure}

For small activity $\zeta$, we report a full phase separation of the density $\phi$ between dilute ($\phi=-1$) and dense ($\phi=1$); the stream function $\psi$ forms either a laminar state [Fig.~\ref{fig:steady_states}(a)] or a vortex state  [Fig.~\ref{fig:steady_states}(b)]. The profiles of $(\phi,\psi)$ have a wavelength of the order to the system size, $L$. The flow speed increases with $\zeta$, particularly between domains of $\phi$ where the flow is concentrated, which leads to destabilizing the laminar flow and favors patterns made of vortices. At even larger $\zeta$, the vortex pattern becomes unstable, and the system enters a self-sustained dynamical steady state [Fig.~\ref{fig:steady_states}(c)], analogous to {\it active turbulence}~\cite{cates2015, Singh2019}. 
In the turbulent state, we measure the velocity spectrum $S(q)$ defined by~\cite{Alert2020}
\begin{equation}\label{eq:S}
	\langle \vb*{v}^2 \rangle = \int q \, \langle |\vb*{v}_{\vb*{q}}|^2 \rangle \,\mathrm{d}q = \int S(q) \,\mathrm{d}q ,
\end{equation}
where $q=|\vb*{q}|$, and $\vb*{v}_{\vb*{q}}$ is the spatial Fourier transform of $\vb*{v}(\vb*{r})$. We observe that $S(q)$ decays monotonically with $q$ [Fig.~\ref{fig:steady_states}(d)]. For AMH with inertia, the scaling $S(q)\sim q^{-7/5}$ has been reported in the turbulent regime~\cite{Bhattacharjee2022}. Here, we cannot infer any specific power-law scalings from our simulations. Numerical measurements for a broader span of $q$ might reveal such scalings. We defer this analysis to future work.


\subsection{Topological defects}\label{sec:topologcal_defects}

The turbulent state is associated with the sustained creation and annihilation of topological defects. We discuss below how these defects can be detected and analyzed as singularities in the profile of the active stress $\Sigma_{\alpha \beta}^{A}$ [Eq.~\eqref{eq:stress_tensor}].

\begin{figure}
    \centering
    \includegraphics[width=.8\linewidth]{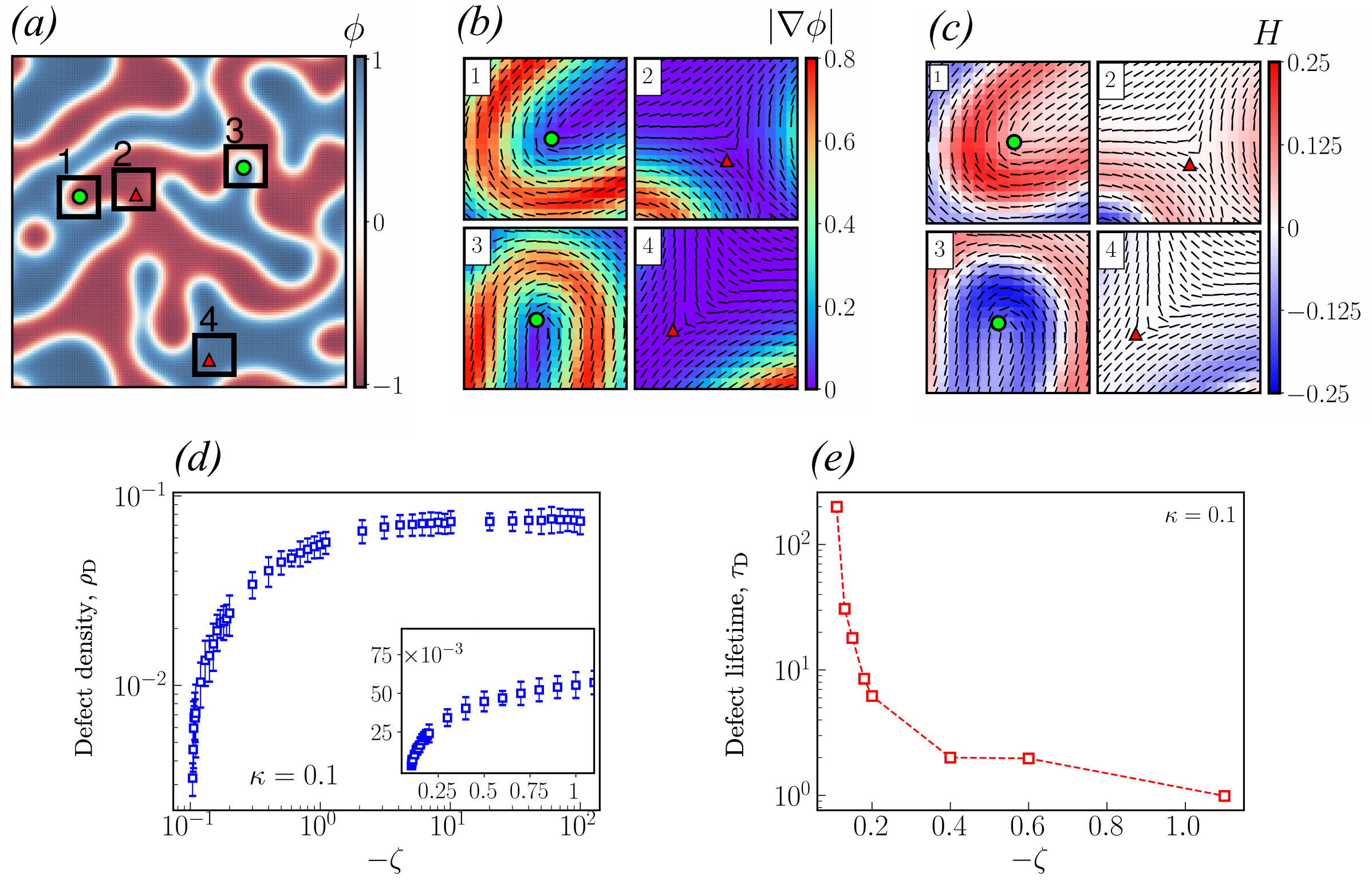}
    \caption{(a)~Density field $\phi$ in the turbulent state {($\zeta=-0.16$)}. The markers indicate the locations of $+\frac 1 2$ (green circles) and $-\frac 1 2$ (red triangles) topological defects. Panels (b,c): zoomed--in view of the four squares marked in panel (a): comparison between the location of the defects, the nematic director field (black lines) along with (b)~the density gradient $|\nabla \phi|$ and (c)~the local curvature $H$.
    (d)~Density of defects $\rho_{\rm D}=\langle N_{\rm D} \rangle/ L^{2}$, where $\langle N_{D} \rangle$ is the defect number averaged over time and realizations, and (e)~defect lifetime $\tau_{\rm D}$, defined as the average time between pair creation and annihilation, as functions of the activity parameter $\zeta$.
    {Parameters $-a=b=0.1$, $\kappa=0.1$, and $\eta=1.67$.}
    }
    \label{fig:gradphi_defect}
\end{figure}

We construct from $\Sigma_{\alpha \beta}^{A}$ a symmetric traceless tensor $Q_{\alpha\beta}=\zeta^{-1}\Sigma_{\alpha \beta}^{A}$ analogous to the nematic order parameter of liquid crystals in 2 dimensions~\cite{DeGennes}:
\begin{equation}
\mathbb{Q} =
	\begin{pmatrix}
	Q_{11} & Q_{12} \\
	Q_{12} & -Q_{11}
	\end{pmatrix} ,
\label{eq:Q}
\end{equation}
where
\begin{equation}\label{eq:q11q12_main}
 	Q_{11} =-\frac{1}{2} \left [ \left ( \partial _{x} \phi\right ) ^{2} -\left ( \partial _{y} \phi\right ) ^{2}\right ] ,
  \quad 
	Q_{12} = - (\partial _{x} \phi) (\partial _{y} \phi) .
\end{equation}
Our interpretation of active stress as a nematic field differs from~\cite{Doostmohammadi2018}, where the nematic field was instead associated with a deformation tensor. We interpret the \emph{orientation} of the principal axes of $\mathbb{Q}$ in analogy with a nematic director. The principal axes are obtained from the eigenvectors $\mathbf{v}_{\pm}$ of $\mathbb{Q}$  by identifying $\mathbf{v}_{\pm} \sim - \mathbf{v}_{\pm}$ [\ref{ap:nematic_director}]. This identification defines a local orientation, which we interpret as the analogue of a nematic director field:
\begin{equation}\label{eq:analogueNematic}
	\mathbf{v}_+ = [(\partial_x \phi)/(\partial_y\phi), 1] \equiv |\mathbf{v}_+|[\cos\theta_N,\sin\theta_N] ,
	\quad
   \theta_N = \tan^{-1}\frac{\partial_y\phi}{\partial_x\phi} . 
\end{equation}
The angle $\theta_N$ represents the orientation of the principal axes of stress if we identify $\theta_N\sim\theta_N+\pi$. In fact, we deduce from Eq.~\eqref{eq:analogueNematic} that $\theta_N$ coincides locally with the direction $\theta$ of $\nabla\phi$. Importantly, while $\theta$ winds from $0$ to $2\pi$ (e.g., $\nabla\phi$ and $-\nabla\phi$ are distinct vectors), the nematic-like orientation is defined modulo $\pi$, since $\mathbf{v}_+$ and $-\mathbf{v}_+$ represent the same orientation. Therefore, the singularities of the nematic tensor $\mathbb{Q}$, namely topological defects where $\theta_N$ is undefined, are located at the critical points of the field $\phi$, namely where $|\nabla\phi|=0$ [Fig.~\ref{fig:gradphi_defect}]. The orientation $\theta_N$ winds  by multiples of $\pi$ around the point-singularities of $\mathbb{Q}$, so the topological charge is quantized in half-integer multiples: $ \frac{1}{2\pi}\oint_\gamma \nabla\theta_N\cdot d\mathbf{\ell}=\frac{s}{2} $, $(s=\pm1,\pm2,...)$ for every loop $\gamma$ enclosing a defect. This connection provides a convenient description to identify and track TDs. {In~\ref{ap:nematic_director}, we show that the amplitude $S$ of the $Q-$tensor reads
\begin{equation}\label{eq:S}
    S = \sqrt{Q_{11}^2+Q_{12}^2} = \frac{1}{2}|\nabla \phi|^2
\end{equation}
which vanishes at the critical points of $\phi$, namely at the center of the TDs, as expected. In active nematics, the core radius is the typical length scale in which $S$ varies from 1 to 0; however, here $S$ is not an order parameter field. The core is set by the typical spatial variation of $|\nabla\phi|$ around the critical points of $\phi$, see for instance Fig.~\ref{fig:gradphi_defect}(b). We refer the reader to \ref{app:bc} for further discussion on the core size.}

In numerical simulations, we compare the orientation of the nematic field $\mathbb{Q}$ and that of $\nabla \phi$ [Figs.~\ref{fig:gradphi_defect}(a-b)]. The locations of topological defects, with charge $\pm \frac 1 2$, indeed coincide with the zeros of $\nabla \phi$. In fact, $+\frac 1 2$ defects are typically located close to interfaces between dense and dilute phases, whereas $-\frac 1 2$ defects are immersed in the bulk phases. It is insightful to interpret the field $\phi(x,y)$ as the height of a topographic surface in the Monge parametrization: this interpretation provides a direct correspondence between the location defects (where $\nabla\phi$ vanishes) and the local curvature $H$ associated with $\phi$ [\ref{ap:nematic_director}]. In fact, the $+\frac 1 2$ defects are located at the maxima and minima of curvature, whereas $-\frac 1 2$ defects are located at the saddle points where $H=0$ [Fig. \ref{fig:gradphi_defect}(c)]. Evaluating the density $\rho_{D}$ [Fig.~\ref{fig:gradphi_defect}(d)] and lifetime $\tau_L$ [Fig.~\ref{fig:gradphi_defect}(e)] of defects, we find that $\rho_\mathrm{D}$ increase monotonically and reaches a plateau at large $\zeta$, whereas $\tau_\mathrm{D}$ decreases monotonically. These measurements support the fact that the reduced lifetime stems from the increase in defect motility at high activity.

\begin{figure}
    \centering
    \includegraphics[width=0.8\linewidth]{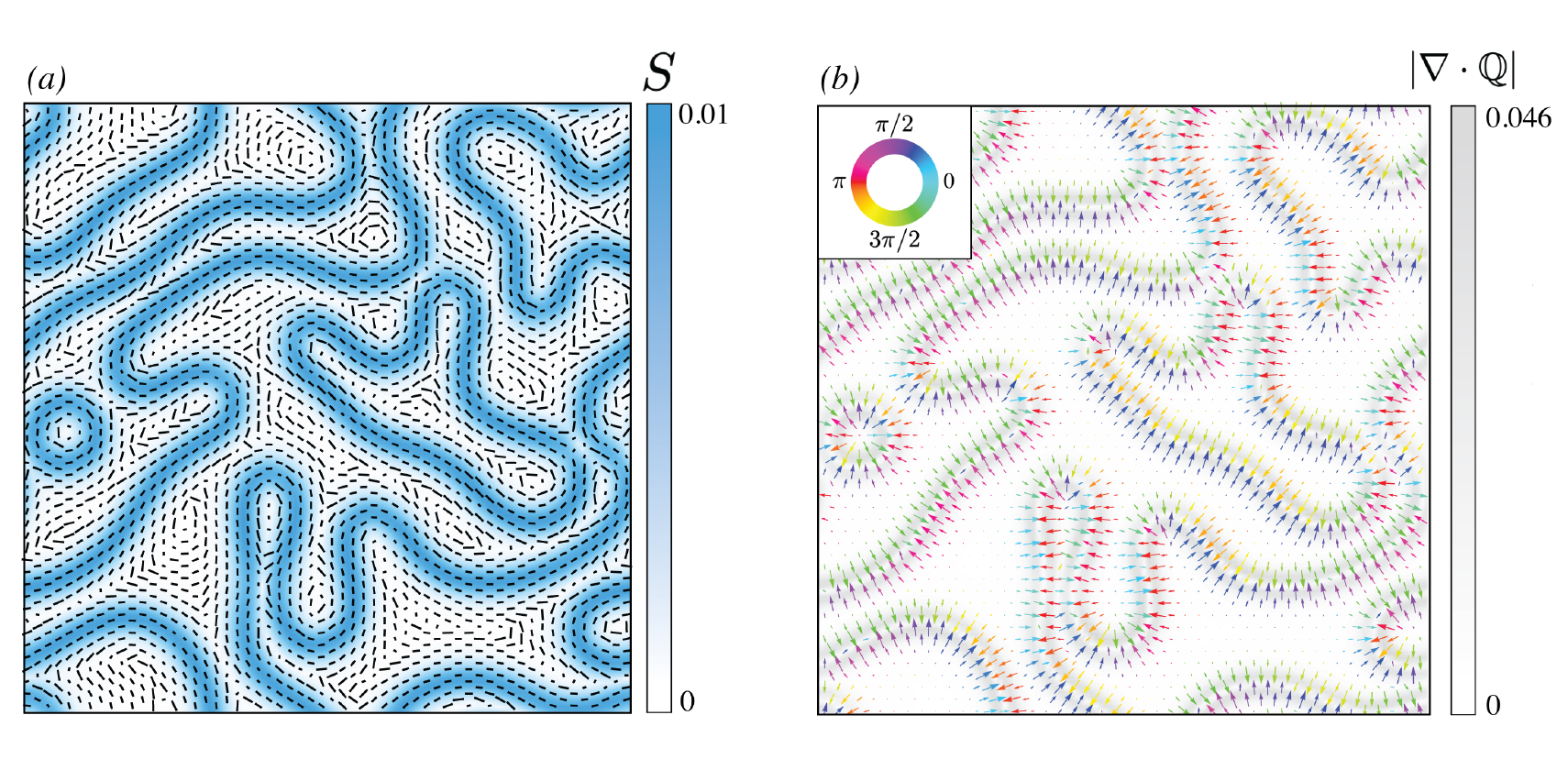}
    \caption{{(a)~Nematic order parameter $\mathbb{Q}$ [Eqs.~\eqref{eq:Q} and~\eqref{eq:q11q12_main}] extracted from the density field $\phi$ in Fig.~\ref{fig:gradphi_defect}(a). The black lines correspond to the director field representing the nematic orientation $\theta_N$ [Eq.~\eqref{eq:analogueNematic}]. The blue colorscale corresponds to the amplitude $S$ of $\mathbb{Q}$ [Eq.~\eqref{eq:S}] that is maximal at interfaces between dense and dilute regions [Fig.~\ref{fig:gradphi_defect}(a)].
    (b)~Total force field is proportional to $\nabla\cdot\mathbb{Q}$ [Eq.~\eqref{eq:fa}]. The gray colorscale corresponds to the amplitude of the total force that is localized at the interfaces. The colored arrows refer to the orientation of the force vectors, see inset. 
}    }
    \label{fig:Q_f}
\end{figure}

In short, the analogy between shear stress and nematic tensor means that singularities in the stress profile must have $\pm \frac 1 2$ charges. Moreover, such defects can actually be detected simply by analyzing the profile of swimmer density $\phi$, without any information on the surrounding flow. {In what follows, to analyze the turbulent phase of AMH, we use three equivalent levels of description: (i)~the density of microswimmers measured by the scalar field $\phi$ [Fig.~\ref{fig:gradphi_defect}(a)], (ii)~the stresses (passive and active) represented by the nematic order parameter $\mathbb{Q}$ [Eq.~\eqref{eq:Q} and Fig.~\ref{fig:Q_f}(a)], and (iii)~the forces (passive and active, see Eq.~\eqref{eq:fa}) which are proportional to $\nabla\cdot\mathbb{Q}$ [Fig.~\ref{fig:Q_f}(b)]. The forces are the sources of the active flows [Eq.~\eqref{eq:stokes}] which stir the topological defects.
}

\section{Quantifying irreversibility from vorticity and topological defects}\label{sec:epr}

In this Section, we propose a quantification of time-reversal symmetry, called informatic entropy production rate (IEPR), and study its features in the turbulent state. We provide a local measure of IEPR and examine its relations with topological defects. We successfully compare our predictions for the local IEPR with simulations in the vicinity of defects.

\subsection{IEPR as a measure of irreversibility}\label{sec:derivation_epr}

Following standard tools of stochastic thermodynamics~\cite{Seifert_2012, Lebowitz_1999, Jack2022}, the IEPR $\dot{\mathcal{S}}$ is defined as
\begin{equation}\label{eq:def_epr}
	\dot{\mathcal{S}}=\lim_{ \tau \to \infty} \frac{1}{\tau} \ln \frac{\mathcal{P}(\{\vb*{J},\psi\}_0 ^{\tau})}{\mathcal{P}^{R}(\{\vb*{J},\psi\}_0 ^{\tau})} \geq 0 ,
\end{equation}
where $\mathcal{P}(\{\vb*{J},\psi\}_0 ^{\tau})$ is the probability of a given trajectory within the time interval $0 \le t \le \tau$, and $\mathcal{P}^R(\{\vb*{J},\psi\}_0 ^{\tau})$ the probability of its time-reversed image. In general, the IEPR depends on the strength of fluctuations. In what follows, we focus on the regime of vanishing noise ($T \to0$) which gives insight into how deterministic dynamical patterns contribute to irreversibility~\cite{nardini_2017, Borthne_2020}. Using field theoretical methods~\cite{janssen_1976, rose_1973, onsager_1953}, we calculate the dominant contribution to IEPR in steady state [\ref{ap:derivation_epr}]:
\begin{equation}\label{loc_epr}
	T \dot{\mathcal{S}} = \int \dot{\sigma} (\vb*{r})  \, \mathrm{d}\vb*{r} ,
	\quad
	\dot{\sigma} (\vb*{r}) = \frac{\eta \zeta}{\kappa+\zeta}\langle \omega^2 \rangle  \geq 0 ,
\end{equation}
where $\omega=-\nabla^2\psi$ is the local vorticity, and $\langle\cdot\rangle$ is an average over time and realizations (under the assumption of ergodicity). Therefore, the IEPR is proportional to the enstrophy $\int \langle \omega^2\rangle \, \mathrm{d}\vb*{r}$~\cite{LandauFluid}, and we have defined the local density of IEPR $\dot{\sigma} (\vb*{r})$. The IEPR vanishes at zero activity ($\zeta=0$), as expected: since activity is the only source of stirring in this system, removing it stops all vortical motions. Note that the regime of zero viscosity is a singular limit, since the assumption of negligible inertia underlying the Stokes equation [Eq.~\eqref{eq:stokes}] breaks down.

\begin{figure}
  \centering
  \includegraphics[width=\linewidth]{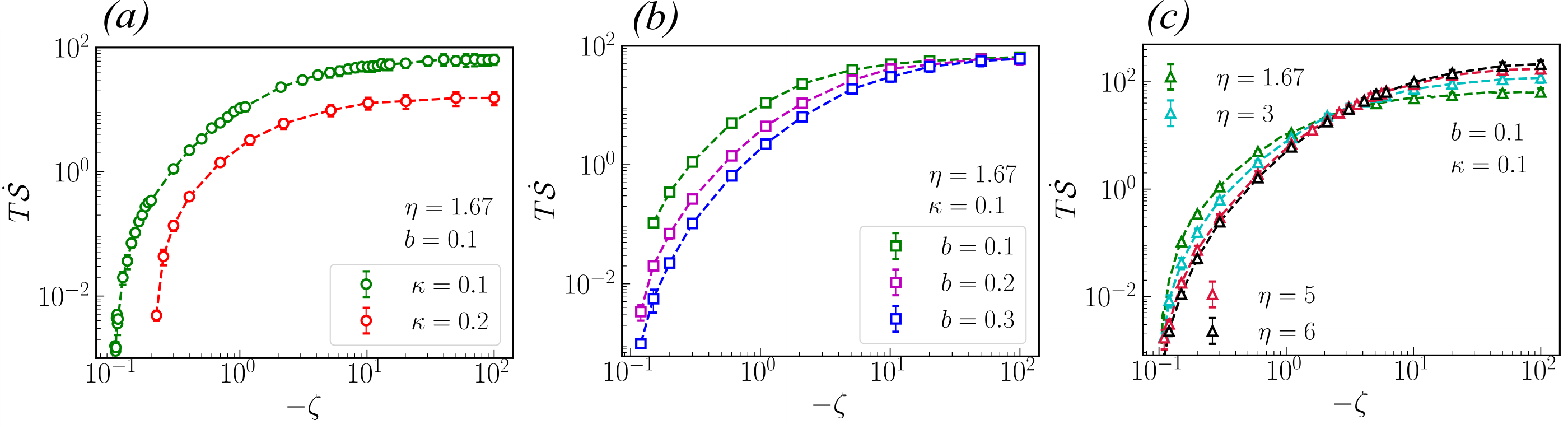}
  \caption{Informatic entropy production rate (IEPR) [Eq.~\eqref{loc_epr}] in steady state as a function of activity $\zeta$, when changing (a)~the cost at density gradients $\kappa$, (b)~the nonlinear free-energy coefficient $b$, and (c)~the viscosity $\eta$. 
  Same parameters as in Fig.~\ref{fig:steady_states}.
  }
  \label{fig:epr_k}
\end{figure}

The local IEPR can also be written in terms of the curl of the velocity field, since $\omega^2=(\nabla\times \vb*{v})^2$:
\begin{equation}
\label{eq:loc_epr2}
   \dot{\sigma} (\vb*{r}) = \frac{\eta \zeta}{\kappa+\zeta} \langle (\nabla\times \vb*{v})^2 \rangle = \frac{\eta \zeta}{\kappa+\zeta} \langle ( \nabla^2 \psi )^{2} \rangle .
\end{equation}
The non-negativity of the local IEPR $\dot{\sigma} (\vb*{r})$ may not be immediately obvious from Eq.~\eqref{loc_epr}. For $-\kappa \leq \zeta \leq 0$, although the term $\zeta/(\kappa+\zeta)$ is negative, the steady state is fully phase-separated state without any flow, so $\dot{\sigma}(\vb*{r})$ vanishes. Note that $\dot\sigma$ is invariant with respect to a gauge transformation $\psi\to\psi+c$ for an arbitrary constant $c$, as required by gauge invariance. In the laminar state $\dot{\mathcal{S}}=0$, since there is no vorticity ($\nabla^2\psi=0$), whereas in the vortex state $\dot{\mathcal{S}}$ does not vanish. In the turbulent state, $\dot{\mathcal{S}}$ first increases rapidly with $\zeta$, and then saturates at large $\zeta$. This observation is robust to variation in $(b,\eta)$ [Fig.~\ref{fig:epr_k}].

In short, Eq.~\eqref{loc_epr} shows that the local IEPR can be extracted from local measurements of the flow field (namely, either $\vb*{v}$, $\omega$ or $\psi$), and that the symmetries of the flow field determine the distribution of IEPR in space. Specifically, the regions of high vorticity $\omega$ are associated with high IEPR $\dot\sigma$. This relation is relevant for estimating the local irreversibility of scalar active matter in experiments where the local flow can be inferred by visual techniques, such as PIV methods~\cite{PIVmethod2021,PIVmethod2018}. 

\subsection{Linear irreversible thermodynamics}\label{sec:LIT}

Some active field theories are postulated from symmetry arguments without any energetic arguments~\cite{Cates2025}, while others are based on the definition of the total dissipation~\cite{Kruse_2004, Marchetti2013}. In many hydrodynamic theories, the dissipation can be written in terms of the products of some thermodynamic forces and currents; linear irreversible thermodynamics (LIT) enforces linear relations between these forces and currents~\cite{Mazur}. To account for the continuous energy injection fueling activity in AMH, we must add to $(\phi,\psi)$ the dynamics of some reactant fields $n$. Inspired by Ref.~\cite{Markovich2021}, we demonstrate how to embed the dynamics of AMH [Eqs.~\eqref{eq:phi_dynamics} and \eqref{eq:stream_dynamics}] in the extended field space $(\phi,\psi,n)$ within the framework of LIT~\cite{Mazur}. Such an embedding provides a route to relating the IEPR [Eq.~\eqref{eq:def_epr}] with the total dissipation.

By analogy with active gel theories~\cite{Kruse_2004}, we assume that our system is in contact with some reservoirs of product and reactant molecules at a fixed chemical potential difference $\Delta\mu$. Molecules undergo chemical reactions at a constant rate $r$, and such reactions generate nonequilibrium mechanical forces that sustain the activity of swimmers. Therefore, to leading order, we assume that the activity parameter $\zeta$ [Eq.~\eqref{eq:stress_tensor}] is proportional to $\Delta\mu$, so we express the active stress $\Sigma^A_{\alpha\beta}$ as follows:
\begin{equation}\label{eq:activeStressDeltamuMain}
	{\Sigma}^A_{\alpha\beta} = - \Delta\mu \,\tilde{\Sigma}^A_{\alpha\beta} \,.
\end{equation}
At small Reynolds number, we write down the Stokes equation [Eq.~\eqref{eq:stokes}] as
\begin{equation}\label{eq:stokes_bis}
	\partial_{\beta} \Sigma^\mathrm{tot}_{\alpha\beta} = 0 ,
\end{equation}
where	
\begin{equation}\label{eq:stress_def}
\begin{aligned}
	\Sigma^\mathrm{tot}_{\alpha\beta} &= 2\eta v_{\alpha\beta} + {\Sigma}^{P,\mathrm{tot}}_{\alpha\beta} - \Delta\mu \tilde{\Sigma}^A_{\alpha\beta} ,
	\\
	{\Sigma}^{P,\mathrm{tot}}_{\alpha\beta} &= {\Sigma}^P_{\alpha\beta} - p\delta_{\alpha\beta} ,
	\\
	2v_{\alpha\beta} &= \partial_\alpha v_\beta+\partial_\beta v_\alpha ,
\end{aligned}
\end{equation}
where $v_{\alpha\beta}$ is the strain rate tensor, and ${\Sigma}^{P,\mathrm{tot}}_{\alpha\beta}$ includes the pressure $p$. We then rewrite the noiseless dynamics of AMH [Eqs.~\eqref{eq:phi_dynamics} and \eqref{eq:stokes_bis}] in terms of the following thermodynamic forces ${\mathbb X}$ and currents ${\mathbb J}$:
\begin{equation}\label{eq:lit}
	{\mathbb X} = \bigg( - \partial_\alpha \frac{\delta\cal F}{\delta\phi}, \,\Sigma^\mathrm{tot}_{\alpha\beta} - {\Sigma}^{P,\mathrm{tot}}_{\alpha\beta} ,\, {\Delta\mu} \bigg)^\mathrm{T} ,
	\quad
	{\mathbb J} = (J_\alpha, v_{\alpha\beta}, r)^\mathrm{T} ,
\end{equation}
that are linearly related through the Onsager matrix $\mathbb L$:
\begin{equation}\label{eq:OnsagerMain}
	{\mathbb J} = {\mathbb L} \cdot {\mathbb X} ,
	\quad
	{\mathbb L} = \begin{pmatrix}
      M & 0 & 0
      \\
      0 & (2\eta)^{-1} & (2\eta)^{-1}\,\tilde{\Sigma}^A_{\alpha\beta} 
      \\
      0 & (2\eta)^{-1}\,\tilde{\Sigma}^A_{\alpha\beta}  &  c_r
  \end{pmatrix}
  .
\end{equation}
In writing Eq.~\eqref{eq:OnsagerMain}, we have introduced the kinetic coefficient $c_r$, and we have used that $\mathbb L$ must be symmetric for dissipative fluxes and forces~\cite{onsager_1953}. The stability of the dynamics enforces that $\mathbb L$ is positive semi-definite~\cite{Mazur}, which provides an explicit constraint on $c_r$ as
\begin{equation}\label{eq:det}
	\int (\det \mathbb L) {\rm d}\vb*{r} = \frac{M}{2\eta} \int \left( c_r - \frac{1}{2\eta} \tilde{\Sigma}^A_{\alpha\beta}\tilde{\Sigma}^A_{\alpha\beta} \right) {\rm d}\vb*{r} \geq 0 .
\end{equation}
As a result, embedding AMH [Eqs.~\eqref{eq:phi_dynamics} and \eqref{eq:stokes_bis}] into the extended field space $(\phi,\psi,n)$ within LIT [Eq.~\eqref{eq:OnsagerMain}] amounts to writing the local chemical rate $r$ as a linear combination of the fluid's velocity $\vb* v$ and the auxiliary field $\tilde\Sigma^A_{\alpha\beta}$:
\begin{equation}\label{eq:r}
	r = c_r \Delta\mu + v_{\alpha\beta} \tilde\Sigma^A_{\alpha\beta} - \frac{\Delta\mu}{2\eta} \tilde\Sigma^A_{\alpha\beta} \tilde \Sigma^A_{\alpha\beta} ,
\end{equation}
where we have used Eq.~\eqref{eq:OnsagerMain}. Note that since the active stress is proportional to $\Delta\mu$ and the latter is constant in space and time, the original AMH dynamics for the swimmer concentration $\phi$ and the stream function $\psi$ remains unchanged.

The rate of energy dissipated by the system can be written as (\cite{Mazur},~\ref{app:LIT})
\begin{equation}\label{eq:diss}
	T\dot{\cal S}_\mathrm{tot} = \int ({\mathbb J} \cdot {\mathbb X}) \mathrm{d} \vb*{r} = - \frac{\mathrm{d}\mathcal{F}}{\mathrm{d}t} + \int ( r \Delta\mu ) \mathrm{d} \vb*{r} \geq 0 .
\end{equation}
Such a linear relation between dissipation, forces, and fluxes holds even in the presence of fluctuations~\cite{Markovich2021}. The term $-{\rm d}{\cal F}/{\rm d}t$ in Eq.~\eqref{eq:diss} accounts for the transient relaxation (and vanishes in steady state), whereas the term $r\Delta\mu$ describes how the profiles $(\phi,\psi)$ affect dissipation. For homogeneous profiles (namely, when $\tilde\Sigma^A_{\alpha\beta}=0$), the IEPR $\dot{\cal S}$ vanishes  [Eq.~\eqref{loc_epr}], whereas substituting the definition of $r$ [Eq.~\eqref{eq:r}] into Eq.~\eqref{eq:diss} yields a positive dissipation rate $\dot{\cal S}_{\rm tot} \sim c_r \Delta\mu^2$. In other words, even in the absence of any patterns in $(\phi,\psi)$, there is a positive dissipation due to the underlying chemical reactions.

For arbitrary profiles $(\phi, \psi)$, substituting the definition of $r$ [Eq.~\eqref{eq:r}] into Eq.~\eqref{eq:diss}, we identify two contributions to the steady-state dissipation $T\dot{\cal S}_\mathrm{tot}^{\,(\mathrm{ss})}$:
\begin{equation}\label{eq:ssepr1}
    \dot{\cal S}^{\,(\mathrm{ss})}_\mathrm{tot} = \frac{(2\eta)^4\Delta\mu ^2}{M}\int ( \det\mathbb{L} )\mathrm{d} \vb*{r} - \int v_{\alpha\beta} \Sigma_{\alpha\beta}^A {\rm d}\vb*{r} .
\end{equation}
We demonstrate in \ref{app:LIT} that the second term in Eq.~\eqref{eq:ssepr1} coincides with the IEPR $T\dot{\cal S}$ [Eq.~\eqref{loc_epr}]. In conclusion, $\dot{\cal S}^{\,(\mathrm{ss})}_\mathrm{tot}$ can be expressed in terms of $\dot{\cal S}$ as
\begin{equation}\label{eq:EPRsplit2Main}
  \dot{\cal S}^{\,(\mathrm{ss})}_\mathrm{tot} = \frac{(2\eta)^4 \Delta\mu^2}{M T} \int  ( \det\mathbb{L} )\mathrm{d} \vb*{r} + \dot{\cal S} .
\end{equation}
Therefore, Eq.~\eqref{eq:EPRsplit2Main} provides an explicit relation between the IEPR of AMH and the steady-state dissipation in the extended hydrodynamic theory. Combining Eqs.~\eqref{eq:det} and~\eqref{eq:EPRsplit2Main}, we deduce that the IEPR gives a lower bound on the total dissipation:
\begin{equation}\label{eq:bound}
	\dot{\cal S}^{\,(\mathrm{ss})}_\mathrm{tot} \geq \dot{\cal S} \geq 0 .
\end{equation}
Defining the chemical concentration $n$ in terms of the chemical rate $r=\partial_t n$, the total dissipation $T\dot{\cal S}_\mathrm{tot}$ measures the irreversibility of the trajectories $(\phi,\psi,n)$~\cite{Markovich2021}. In contrast, $\dot{\cal S}$ only accounts for the irreversibility of $(\phi,\psi)$, so it provides only a partial estimation of the total dissipation. In short, the bound in Eq.~\eqref{eq:bound} illustrates how irreversibility decreases when considering a reduced set of dynamical fields~\cite{Esposito2012}.


\subsection{Local measurement of irreversibility: the role of topological defects}\label{sec:epr_defects}

\begin{figure}
    \centering
    \includegraphics[width=1\textwidth]{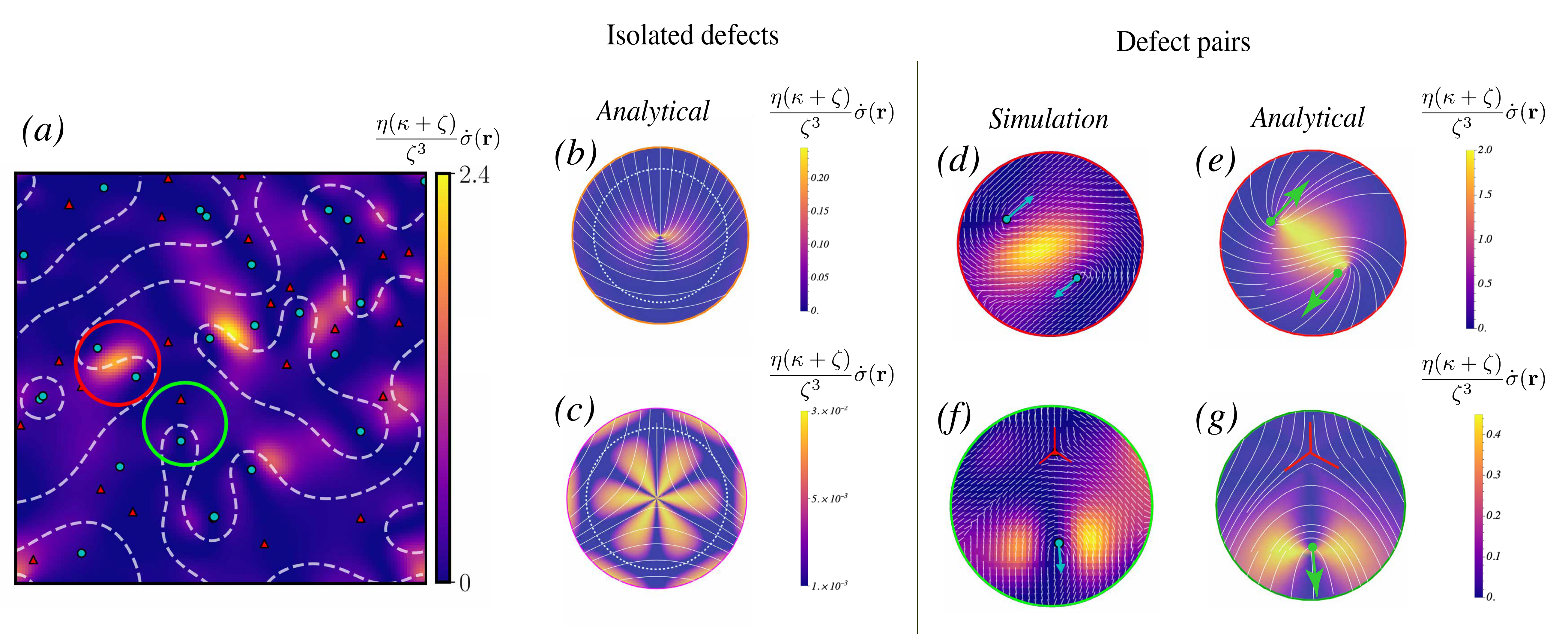}
    \caption{{(a)~Spatial distribution of the local IEPR $\dot\sigma$ [Eq.~\eqref{eq:loc_epr2}] extracted from the same turbulent state as in Figs.~\ref{fig:steady_states}(c) and~\ref{fig:gradphi_defect}(a). The white dashed lines represent the phase boundary $\phi =0$. The green dots and red triangles respectively refer to $+\frac{1}{2}$ and $-\frac{1}{2}$ topological defects.
    (b,c)~Analytic predictions of $\dot\sigma$ around isolated topological defects [Eq.~\eqref{eq:sf1}] of charge $+\frac{1}{2}$ [panel~(b)] and $-\frac{1}{2}$ [panel~(c)]. White lines refer to the nematic orientation.
    (d-g)~Comparison between simulations and analytical predictions for defect pairs. A pair with charges $(+\frac{1}{2},+\frac{1}{2})$~(d,e) and a pair with charges $(+\frac{1}{2},-\frac{1}{2})$ (f,g) marked respectively with red and green circles in panel (a).  We numerically evaluate the relative angle $\Phi$ and the distance $d$ between defects [Sec.~\ref{sec:pair}]. White lines refer to the nematic orientation.}
    }
  \label{fig:epr}
\end{figure}

{In the turbulent state, the local IEPR $\dot{\sigma}$ fluctuates in the form of sharply localized regions that continuously appear and disappear in time; we refer the reader to the Supplementary video. We report in Fig.~\ref{fig:epr}(a) a typical instantaneous configuration, where the field $\dot{\sigma}$ is directly evaluated from $\psi$ [Eq.~\eqref{eq:loc_epr2}], showing that $\dot{\sigma}$ is typically larger at interfaces (between small- and large-$\phi$ regions) than in the bulks, and also higher in regions of large vorticity. We observe two necessary conditions for regions of high IEPR to emerge: (i)~the interfaces form S-shaped meandering curves, and (ii)~the curved interfaces enclose pairs of $+\frac{1}{2}$ TDs which are oriented perpendicular to the line connecting them [red circle in Fig.~\ref{fig:epr}(a) and Figs.~\ref{fig:epr}(d,e)]. In general, other configurations of paired TDs can produce a high IEPR locally: in a pair of $+\frac{1}{2}$ and $-\frac{1}{2}$ TDs, $\dot\sigma$ is locally higher close to the $+\frac{1}{2}$ defect [Fig.~\ref{fig:epr}(f)]. 
}

{In what follows, we provide analytical predictions that reproduce the IEPR profiles observed in the simulations, and argue that such profiles can be rationalized in terms of the symmetries of the TDs in the nematic field $Q$ [Eq.~\eqref{eq:Q}]. To this end, we examine \textit{idealized} TDs by considering either isolated TDs of charge $+\frac{1}{2}$ and $-\frac{1}{2}$ [Sec.~\ref{sec:isolatedDefects}], or pairs of TDs obtained by superposing the orientation fields of two isolated defects [Sec.~\ref{sec:pair}].}

\subsubsection{Isolated defects.}\label{sec:isolatedDefects}

{We predict analytically the intensity and the spatial symmetries of the local IEPR $\dot\sigma$ around isolated defects using Eq.~\eqref{eq:loc_epr2}.} The flow field around isolated defects can be found analytically by assuming that {(i)~the amplitude of the nematic $Q-$tensor is constant, and (ii)~}the density field $\phi$, which determines the local stress [Eq.~\eqref{eq:stress_tensor}], relaxes faster than the stream function $\psi$, which determines the local flow. The stream function $\psi$ associated with the flow field $\vb*{v}$ of an isolated defect was derived in Ref.~\cite{Giomi2014}. Using polar coordinates $\vb*{r}=(r,\varphi)$ centered at the defect core, the solution can be written as $\psi=\bar\psi+\psi_0$, where
\begin{equation}\label{eq:sf1}
	\bar\psi_{+ \frac{1}{2}}(\vb*{r}) = \frac{\zeta}{12\eta}\, r (3R-2r) \sin \varphi ,
	\quad
	\bar\psi_{- \frac{1}{2}}(\vb*{r}) = \frac{\zeta}{240 \eta R} r^2 (5 r - 8 R)\sin 3 \varphi ,
\end{equation}
and the term $\psi_0=(a_1r+b_1r^3)\sin\varphi$ is a solution to the homogeneous the Stokes equation, with $(a_1,b_1)$ being fixed by the boundary conditions. Following~\cite{Giomi2014}, we assume that there is a disc of radius $R$ centered at the isolated defect which does not contain other defects. We impose vanishing radial flux through the disc's boundary ($v_r(R,\varphi)=0$), and force-balance tangentially to the boundary: $v_\varphi(R,\varphi)=-\xi \Sigma^{A}_{r\varphi}$ where $\xi\sim \eta h$ and $h$ is the thickness of the boundary layer between neighboring defects~\cite{Giomi2014}. The full solution then reads
\begin{subequations}\label{eq:sf1}
\begin{align}
\label{eq:sfPlus}
  \psi_{+ \frac{1}{2}}(\vb*{r}) &= \frac{\zeta}{2\eta}\left[ \frac{1}{2} \biggl( \frac{R}{3} + \frac{\eta}{\xi}\biggl)r -\frac{1}{3}r^2 + \frac{1}{2} \biggl( \frac{R}{3} - \frac{\eta}{\xi}\biggl) r^3  \right]\sin\varphi ,
  \\
  \psi_{- \frac{1}{2}}(\vb*{r}) &= -\frac{\zeta}{2\eta}\frac{r^2\left(2 r^3 (15 \eta +\xi  R)-r R^2 (\xi  R+30 \eta )+4 \xi  R^4\right)}{60   \xi  R^4} \sin (3 \varphi ) .
\end{align}
\end{subequations}
Substituting the profiles Eq.~\eqref{eq:sf1} into the expression of the local IEPR [Eq.~\eqref{loc_epr}], we deduce
\begin{subequations}\label{eq:sig}
\begin{align}
    \label{eq:eprPlusHalf}
	T\dot{\sigma} _{+\frac{1}{2}}(\vb*{r}) & = \frac{\zeta ^{3} }{\eta(\zeta+\kappa)} \biggl[\frac{1}{4} + \biggl(\frac{\eta}{R\xi }-\frac{1}{3}\biggl) \frac{r}{R}  \biggl]^2 \sin ^{2}\varphi ,
	\\
    \label{eq:eprMinusHalf}
	T\dot{\sigma} _{-\frac{1}{2}}(\vb*{r}) & =  \frac{\zeta ^{3} }{\eta(\zeta+\kappa)}\left[ \frac{1}{6}  -  4\left(\frac{\eta}{R\xi }  + \frac{1}{15}\right) \left(\frac{r}{R} \right)^3  \right]^2\sin ^2(3 \varphi ) .
    \end{align}
\end{subequations}
{The spatial symmetries of the stream function are reflected in the profile of the IEPR [Figs.~\ref{fig:epr}(b,c)].} Our prediction of $\dot\sigma$ [Eq.~\eqref{eq:sig}] assume no-slip boundary conditions, and is qualitatively independent of the choice of boundary conditions close to the defect core (i.e. far from the boundary layer). A different choice would only change our prediction quantitatively while preserving the angular symmetries~\cite{Giomi2014}. In \ref{app:bc}, we demonstrate the existence of an inner region ($r < r_{C}$) where the IEPR is insensitive to the boundary conditions for all isolated $\pm \frac 1 2$ defects.


In short, the local IEPR around isolated defects is spatially inhomogeneous, with symmetries stemming from the vorticity of the backflow around defects: for the $+\frac 1 2$ defect, it is mirror-symmetric across the direction of the polarization; for the $-\frac 1 2$ defect, it features a six-fold rotational symmetry. The proximity of defects of opposite charges induces perturbations of these profiles. Generally, perturbations of the IEPR profile are more pronounced for $-\frac 1 2$ defects. {In typical configurations [Fig.~\ref{fig:epr}(a)],  there are no rigorously isolated defects due to the highly correlated distortions of the interfaces across the sample.}

{The assumption of a constant amplitude of the $Q-$tensor that underlies our derivation leads to quantitative differences between the analytic and the simulated profiles. In simulations, the active force $f^A_\alpha\propto \partial_\beta Q_{\alpha\beta}$ [Eq.~\eqref{eq:fa}] around $\pm \frac{1}{2}$ defects is localized along the interfaces of $\phi$, and vanishes at the center of the defects, whereas it is maximal at the center in our analytic solution. The analysis and justification of these discrepancies are deferred to \ref{app:active_force}.
}


\subsubsection{Defect pairs.}\label{sec:pair}

To predict the profiles of the local IEPR $\dot\sigma$ around defect pairs, we construct the director field $\theta_N(\vb*{r})$ by superposing the textures of isolated defects. For a pair of defects with charges $(q_{1},q_{2})$, the configuration is described by two parameters: the separation $d$ between the defect cores, and the relative angle $\Phi$ between the defect orientations. The orientation of the nematic director field [Eq.~\eqref{eq:Q}] reads
\begin{equation}\label{eq:thetaPair}
	{\theta}_N(x,y) = q_{1} \arctan\left (\frac{y\cos \Phi + \left ( x+d/2 \right) \sin \Phi}{\left ( x+d/2 \right) \cos \Phi-y\sin \Phi} \right )+q_{2} \arctan\left (\frac{y}{x-d/2} \right ) .
\end{equation}
{We assume that the scale of the defect cores is small compared to the typical variations of the IEPR profile, and that the amplitude} $S=\sqrt{Q_{11}^2 + Q_{12}^2}$ is constant outside the defect cores, so that the variations of $\mathbb{Q}$ are entirely determined by the orientation $\theta_N$. The velocity field $\vb*{v}$ associated with the backflow of the defect pair follows by solving the Stokes equation [Eq.~\eqref{eq:stokes}] {by using a Green function: $v_{\alpha}(\vb*{r}) =  \int \mathrm{d}^2\vb*{r}{'}G_{\alpha \beta}(\vb*{r}-\vb*{r}')  f_{\beta}(\vb*{r}') $ where $f$ is the force acting on the fluid. Using $f_\beta=\partial_\gamma \Sigma_{\beta\gamma}$ we can write:}
\begin{equation}\label{eq:134}
	v_{\alpha}(\vb*{r}) = {(\kappa + \zeta)} \int \mathrm{d}^2\vb*{r}{'}G_{\alpha \beta}(\vb*{r}-\vb*{r}') \partial_\gamma Q_{\beta\gamma}(\vb*{r}') ,
\end{equation}
where the Oseen tensor $\mathbb{G}$ in two dimensions reads~\cite{Di_Leonardo_2008}:
\begin{equation}\label{eq:Oseen}
	G_{\alpha\beta}(\vb*{r})=\frac{1}{4 \pi \eta}\left [ \left ( \log \frac{\mathcal{L}}{r}-1\right ) \delta_{\alpha\beta} + \frac{r_{\alpha}r_{\beta}}{r^{2}} \right ] .
\end{equation}
The parameter $\mathcal{L}=R/\sqrt{e}$ is relevant for no-slip boundary conditions. Note that the divergence of $G_{\alpha\beta}$ as $r\to 0$ is consistent with the breakdown of a continuum description near the defects' core.

Finally, the local IEPR follows from substituting the solution for the flow [Eq.~\eqref{eq:134}] into Eq.~\eqref{loc_epr}. Below, we examine three cases at a fixed distance ($d=R/2$) when changing the orientation $\Phi$. {The orientation of the anisotropic $+\frac 1 2$ defects is described by the polarization vector defined as $p_\alpha = - \partial_\beta Q_{\alpha\beta} (\vb*{r}=\vb*{r}_+)$, where $\mathbb{Q}$ is the nematic field [Eq.~\eqref{eq:Q}] and $\vb*{r}_+$ is the position of the defect core~\cite{Shankar_2019}.} For all profiles, the local vorticity changes as $\omega \to - \omega$ under the transformation $\Phi\to\Phi+\pi$; given that $\dot{\sigma}$ is quadratic in $\omega$ [Eq.~\eqref{loc_epr}], it is invariant under such a transformation, so we consider only the range $0\leq \Phi \leq \pi/2$.

\begin{figure}
	\centering
	\includegraphics[width=.8\linewidth]{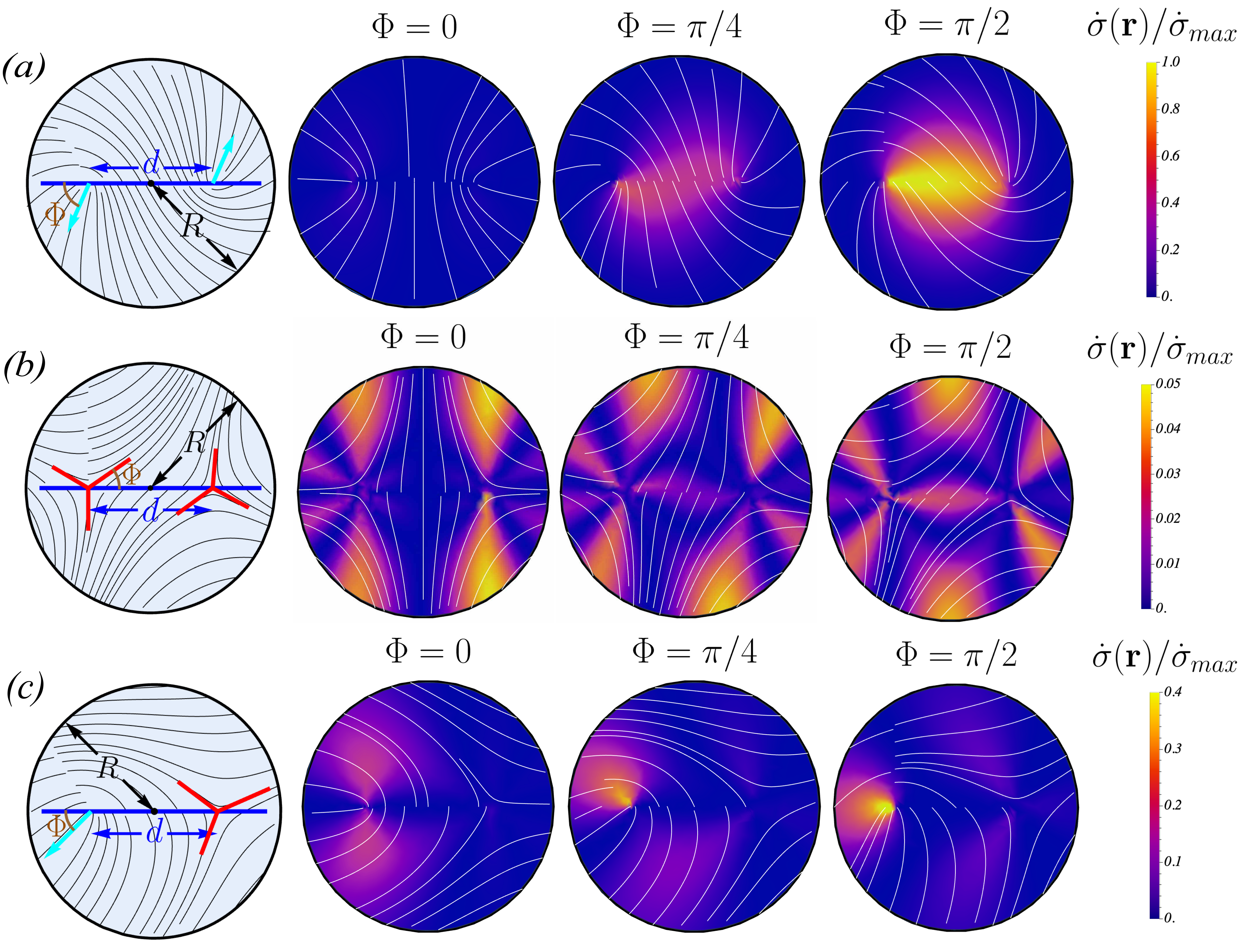}
	\caption{Analytical prediction for the local IEPR $\dot{\sigma}$ [Eq.~\eqref{loc_epr}]: (a) a pair of $+\frac{1}{2}$ defects, (b) a pair of $-\frac{1}{2}$ defects, and (c) a pair of $+ \frac{1}{2}$ and $- \frac{1}{2}$ defects. The defects are at a fixed distance $d$ within a disc of radius $R$, and the angle of orientation changes from $\Phi=0$ to $\Phi=\pi/2$. The first panel in each row shows the schematic of the configuration. For all panels, the color bars are normalized by the maximum value $\dot{\sigma}_{max}$ attained for a pair of $+\frac{1}{2}$ defects within the disc.
	}
	\label{fig:epr_def_orientation}
\end{figure}

Let us start with a pair of $+\frac{1}{2}$ defects  [Fig.~\ref{fig:epr_def_orientation}(a)]. At $\Phi=0$, the defects have collinear and opposite polarizations that are parallel to the segment that connects the cores; the vortices around each defect interfere negatively, yielding low $\dot{\sigma}$. At $\Phi=\pi/2$, the polarizations are opposite and perpendicular to the connecting segment; the vortices around each defect coalesce into a large vortex, yielding a high $\dot{\sigma}$. In contrast, a $-\frac{1}{2}$ pair does not yield large $\dot{\sigma}$ as we rotate defects [Fig.~\ref{fig:epr_def_orientation}(b)], since vortices do not coalesce as for a $+\frac 1 2$ pair. For a $(+\frac 1 2,-\frac 1 2)$ pair [Fig.~\ref{fig:epr_def_orientation}(c)], $\dot{\sigma}$ is highest at $\Phi=\pi/2$ as the left vortex of the $+\frac{1}{2}$ does not interfere negatively with the vortices of the $-\frac 1 2$ defect. In short, the vortices generated by the total force regulate the localization of IEPR. Since the total force is localized along the interfaces between regions of small and large $\phi$ [Fig.~\ref{fig:Q_f}(b)], it follows that $\dot\sigma$ is also highest at interfaces.


We compare our analytical predictions with some defect pairs spontaneously formed in the turbulent state. For a $+\frac 1 2$ pair where $(d,\Phi)\simeq(0.45 R, 0.45\pi)$ [Fig.~\ref{fig:epr}(e)], $\dot{\sigma}$ is higher between defects, in agreement with our predictions [Fig.~\ref{fig:epr_def_orientation}(a)]. For a $(+\frac 1 2,-\frac 12)$ pair where $(d,\Phi)\simeq(0.49 R, 0.95 \pi)$ [Fig.~\ref{fig:epr}(g)], $\dot{\sigma}$ is higher close to the $+\frac 1 2$ defect, and symmetric with respect to its polarization orientation, in agreement with our predictions [Fig.~\ref{fig:epr_def_orientation}(c)]. In both cases, we attribute the slight quantitative deviations to the spatial asymmetries in the numerical solution of the nematic field.

\begin{figure}
	\centering
	\includegraphics[width=.8\linewidth]{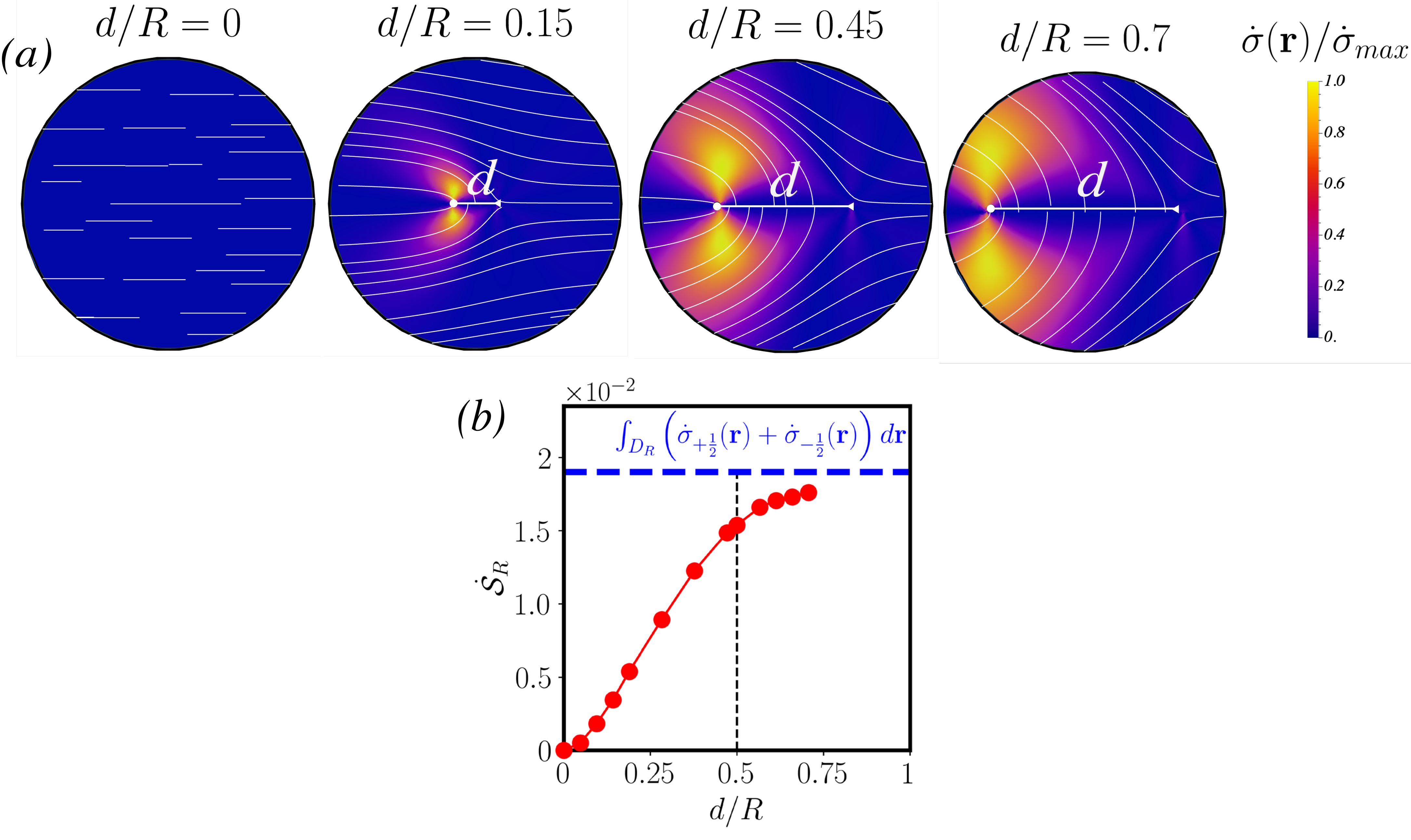}
	\caption{Analytical prediction for the local IEPR $\dot{\sigma}$ [Eq.~\eqref{loc_epr}] as a function of the distance $d$ between defects within a disc of radius $R$. (a)~Series of snapshots of $\dot{\sigma}$ for varous values of $d$. The colorbars are normalized by the maximum $\dot{\sigma}_{max}$ attained within the disc. (b)~The integrated IEPR $\dot{\cal S}_{R}$ [Eq.~\eqref{int_epr}] as a function of $d/R$. The blue dashed line is the sum of the IEPR for the isolated $+\frac{1}{2}$ and $-\frac{1}{2}$ defects in a disc of the same radius $R$.
	}
	\label{fig:epr_creation}
\end{figure}

Finally, we consider a symmetric configuration of a $(+\frac 1 2,-\frac 1 2)$ pair [Fig.~\ref{fig:epr_creation}(a)], and examine the effect of parametrically varying the separation distance $d$ at a fixed orientation $\Phi=0$. For each value of $d$, we introduce the integrated IEPR as
\begin{equation}\label{int_epr}
    \dot{\cal S}_R = \int_{D_R}\dot{\sigma}(\vb*{r})\, \mathrm{d}^2 \vb*{r} ,
\end{equation}
where $D_R$ is the disc of radius $R$ centered at the mid-point between defect cores. We observe that $\dot{\cal S}_R$ increases monotonically with $d$: it vanishes at $d=0$, and approaches the sum of contributions from isolated defects as $d$ increases, as expected [Fig.~\ref{fig:epr_creation}(b)]. In the turbulent state, the fluid is continuously stirred by the creation and annihilation of $(+\frac 1 2,-\frac 1 2)$~\cite{Giomi_2013}. It is tempting to interpret our symmetric configurations [Fig.~\ref{fig:epr_creation}(a)] as a time-series of snapshots during such a creation (namely, for increasing $d$) and annihilation (namely, for decreasing $d$). Our results suggest that the IEPR increases/decreases monotonically when two defects are brought further/closer infinitely slowly.


\section{Discussion}

We have examined the relation between the spatial distribution of the informatic entropy production rate (IEPR)~\cite{Jack2022} and the location of topological defects~\cite{Giomi_2013} in active turbulence~\cite{Alert2022}. Based on a minimal hydrodynamic theory, coupling the density of swimmers and the stream function of the surrounding fluid, we have performed numerical simulations and studied the self-sustained flows that emerge spontaneously despite the absence of inertia. Chaotic (turbulent) flows emerging in active model H were reported in \cite{cates2015}. In this work, we observed two novel flow states in active model H, which we called the laminar and vortex states. This result extends the portfolio of inertia-less turbulence to active model H~\cite{cates2015}, which goes beyond the well-studied case of active nematics~\cite{Alert2020}.

Using the formalism of stochastic thermodynamics~\cite{Seifert_2012, Lebowitz_1999, Jack2022}, we demonstrate that the dominant contribution to IEPR is generically located in regions of high vorticity. This contribution is due to the coupling between the momentum-conserving fluid (governed by the Stokes equation) and the density of swimmers through active stress. Moreover, using linear irreversible thermodynamics, we have shown that the IEPR represents only a partial estimation of the total dissipation rate. To analyze in detail the spatial profile of IEPR, we build a mapping between the shear stress, which stirs the fluid, and an effective nematic field, by analogy with liquid crystals~\cite{Alert2020}. Our mapping leads to identifying some $\pm\frac 1 2$ defects, with symmetries analogous to that of nematics, defined as locations where the gradient of the swimmer density vanishes. Similarly to active nematics, topological defects stir the flow and shape the spatial structure of vortices. Here, we have shown that, remarkably, the vortices near {pairs of} topological defects indicate the dominant contribution to irreversibility. Therefore, our mapping reveals a direct relation between topological defects and IEPR.

We provide analytical derivations for the profile of IEPR around $+\frac 1 2$ and $-\frac 1 2$ defects. Our results predict that the dominant contribution to IEPR stems from configurations where nearby $+\frac{1}{2}$ defects are oriented orthogonally to each other, in direct agreement with numerical simulations. Moreover, we examine the dependence of IEPR on the distance between $(+\frac 1 2,-\frac 1 2)$ pairs, as a first step towards quantifying IEPR during the creation/annihilation of such pairs (a detailed analysis requires resolving the TDs dynamics during creation/annihilation)~\cite{Pearce2025}. {We show that the main features of the local IEPR can be predicted analytically by considering point-like idealized TDs, both isolated and paired. Taking into account a finite core size for TDs, we can even predict the finer details of the local IEPR.} 
These results open some unprecedented perspectives for quantifying irreversibility in experimental realizations of active turbulence. Moreover, the strong localization of IEPR in small regions centered around the point-like TDs indicates that the total IEPR across the system may be estimated by a discrete number of local measurements. It would be interesting to explore further whether the fluctuations of IEPR are also constrained by the TDs statistics~\cite{Mahault2022}.

Recently, some protocols have been proposed to control the dynamics of flows and defects in active nematics~\cite{Shankar2024, Hagan2025} in line with experiments~\cite{Gardel2021, Fraden2024}. From a broader perspective, different control strategies have been put forward to stabilize patterns in various active systems~\cite{Shankar2022, Davis2024, alvarado2025}. To this end, an important challenge is to identify the relevant degrees of freedom which determine the emerging dynamical patterns. By analyzing how the defect statistics affects the IEPR, we provide the first step towards building protocols which drive the dynamics towards target states at a minimal cost. Our results motivate the search for generic control principles: how local perturbation can result in optimally controlling the dynamics of complex systems~\cite{EPN2024}.


\section*{Acknowledgements}

It is a pleasure to acknowledge fruitful discussions with Sumesh P. Thampi. \'EF acknowledges discussions with Michael E. Cates and Rajesh Singh. This research was funded in part by the grant reference 14389168 (Luxembourg National Research Fund, FNR), and by the grant no. NSF PHY-2309135 to the Kavli Institute for Theoretical Physics (KITP). B.N.R. acknowledges support from the FNR grant PRIDE19/14063202/ACTIVE.


\appendix

\section{Stochastic dynamics of the stream function}\label{ap:stream_derivation}

To derive the stochastic dynamics of the stream function [Eq.~\ref{eq:stream_dynamics}], we apply the curl operator to the Stokes equation [Eq.~\ref{eq:stokes}]. The term that contains the gradient of pressure vanishes:
\begin{equation}\label{eq:pressure}
    \nabla \times (\nabla p)=0 ,
\end{equation}
and the viscous term gives
\begin{equation}
	\nabla \times (\eta \nabla^{2}\mathbf{v}) = -\eta\nabla^{4}\psi ,
\end{equation}
where $\nabla \times$ is the curl in 2 dimensions and we have used the fact that curl and Laplacian operators commute, and $\partial_{y}v_{x}-\partial_{x}v_{y} = \nabla^{2}\psi$ by definition of the stream function [Eq.~\eqref{eq:streamF}]. Defining the force $f$ as
\begin{equation}\label{eq:force}
	f_{\alpha} = \partial _{\beta}( \Sigma_{\alpha \beta}^{P} + \Sigma_{\alpha \beta}^{A}) = (\zeta+\kappa) \big[ (\delta_{\alpha \beta}/2) \partial _{\beta}(\partial_{\gamma\gamma}\phi)^{2}-\partial _{\beta}(\partial_{\alpha}\phi\partial_{\beta}\phi) \big] ,
\end{equation}
where we have used the definitions of $\Sigma_{\alpha \beta}^{P}$ and $\Sigma_{\alpha \beta}^{A}$ [Eq.~\eqref{eq:stress_tensor}], we get
\begin{equation}
\begin{aligned}
	\frac{f_{x}}{\kappa+\zeta} &= \frac{1}{2}\partial _{x} \left [(\partial_{y}\phi)^{2}-(\partial_{x}\phi)^{2} \right]-\partial _{y}\big[(\partial_{x}\phi)(\partial_{y}\phi) \big] ,
	\\
	\frac{f_{y}}{\kappa+\zeta} &= \frac{1}{2}\partial _{y}\left [(\partial_{x}\phi)^{2}-(\partial_{y}\phi)^{2} \right ] - \partial _{x} \big[(\partial_{x}\phi)(\partial_{y}\phi) \big] ,
\end{aligned}
\end{equation}
from which we deduce
\begin{equation}\label{eq:force_bis}
\begin{aligned}
	\frac{\partial_{x}f_{y} - \partial_{y}f_{x}}{\kappa+\zeta} &= (\partial_{x}\phi) \left[ -\partial_{x}(\partial_{xy}^{2}\phi)+\partial_{y}(\partial_{yy}^{2}\phi)+2\partial_{x}(\partial_{xy}^{2}\phi) \right] 
	\\
	&\quad +(\partial_{y}\phi) \left[ -\partial_{x}(\partial_{xx}^{2}\phi)+\partial_{yy}^{2}(\partial_{x}\phi)-2\partial_{x}(\partial_{yy}^{2}\phi) \right]
	\\
	&=	(\partial_{x}\phi)\nabla^{2}(\partial_{y}\phi) -(\partial_{y}\phi)\nabla^{2}(\partial_{x}\phi) .
\end{aligned}
\end{equation}
Applying the curl operator $\epsilon_{\alpha \alpha'}\partial_{\alpha'}$ to the noise term $\partial_\beta\Gamma_{\alpha \beta}$, where $\epsilon$ is the Levi-Civita matrix ($\epsilon_{xy}=-\epsilon_{yx}=1)$, the noise term $\Xi=\epsilon_{\alpha \alpha'}\partial_{\alpha'\beta}^{2}\Gamma_{\alpha \beta}$ has Gaussian statistics with zero mean and correlation given by 
\begin{equation}\label{eq:psi_noise}
\begin{aligned}
	\langle \Xi(\vb*{r},t) \Xi(\vb*{r}',t') \rangle &= \epsilon_{\alpha \alpha'} \epsilon_{\mu \mu'} \left ( \delta_{\alpha \mu}\delta_{\beta \nu} +\delta_{\alpha \nu} \delta_{\beta \mu} \right ) \partial_{\alpha'\beta \mu' \nu }^{4} \delta(\vb*{r}-\vb*{r}')\delta(t-t')
	\\
	&= \left ( \epsilon_{\alpha \alpha'} \epsilon_{\alpha \mu'} \partial_{\alpha' \mu' }^{2}\nabla^{2} + \epsilon_{\alpha \alpha'} \epsilon_{\mu \mu'}\partial_{\alpha \alpha'\mu \mu' }^{4} \right ) \delta(\vb*{r}-\vb*{r}')\delta(t-t')
	\\
	&= \nabla^{4}\delta(\vb*{r}-\vb*{r}')\delta(t-t') ,
	\end{aligned}
\end{equation}
where we used the identities $\epsilon_{\alpha \alpha'}\epsilon_{\alpha \mu'}=\delta_{\alpha'\mu'}$ and $\epsilon_{\alpha \alpha'}\epsilon_{\mu \mu'}=\delta_{\alpha\mu}\delta_{\alpha'\mu'}-\delta_{\alpha\alpha'}\delta_{\mu\mu'}$. Therefore, combining Eqs.~(\ref{eq:pressure}-\ref{eq:psi_noise}), we deduce the stream {function's} dynamics in Eq.~\eqref{eq:stream_dynamics}.


\section{Numerical methods}\label{ap:numerical_method}

To integrate the hydrodynamic equations [Eqs.~\ref{eq:phi_dynamics} and \ref{eq:stream_dynamics}], we use a pseudo-spectral numerical scheme. The simulations are performed in a square lattice of size $L \times L$, and the space is discretized as $L=N \Delta$. The grid spacing $\Delta$ is kept constant while varying the $N$ to change the system size. The initial condition for $\phi$ is sampled from randomly in $[-0.1,0.1]$, and $\psi$ is set to zero.

\begin{figure}
    \centering
    \includegraphics[width=\linewidth]{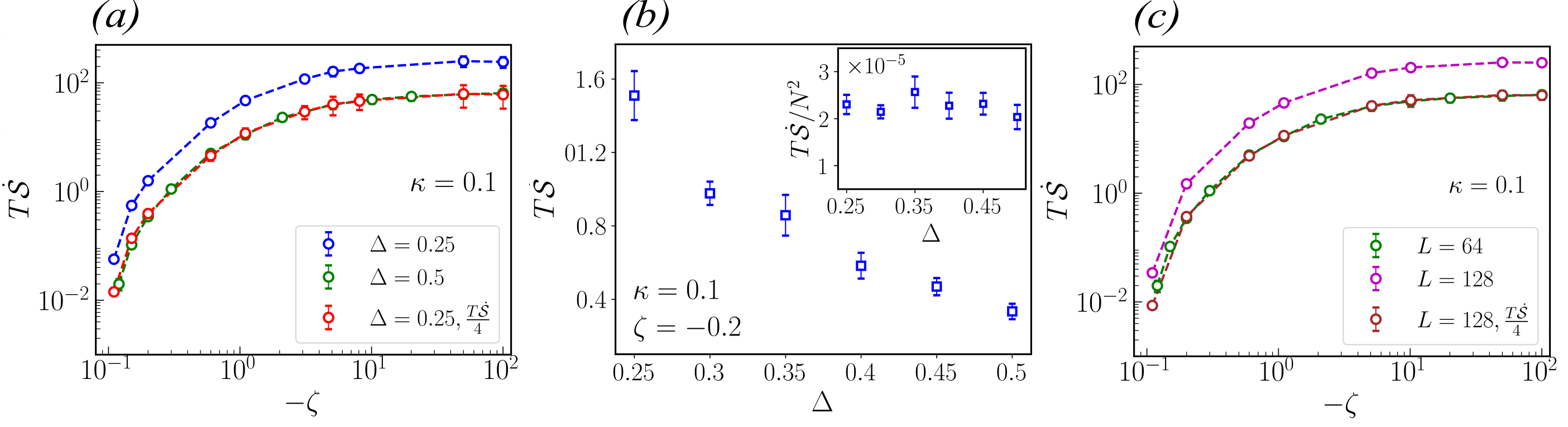}
    \caption{(a)~EPR is plotted as a function of activity for various spatial discretizations $\Delta$ by fixing $L$. (b)~Variation of EPR with spatial discretizations. The inset plot shows EPR per lattice cell. (c)~Plot is same as (a) but by changing $L$ with fixed $\Delta$.}
    \label{fig:epr_dx}
\end{figure}

At each time step, the numerical scheme computes the Fourier transform of density ${\bf F}[\phi(\vb*{r},t)] = \phi_{\vb*{q}}(t)$ and deduces $\psi_{\vb*{q}}$ as
\begin{equation}
	\eta \vb*{q}^{4} \psi_{\vb*{q}} = (\kappa+\zeta) \mathbf{F} \big[ (\partial_x \phi)\nabla ^2 (\partial_y \phi) - (\partial_y \phi)\nabla ^2 (\partial_x \phi) \big] + \sqrt{2\eta T}\vb*{q} ^2 \Lambda_{\psi \vb*{q}} ,
\end{equation}
where we have used Eq.~\eqref{eq:stream_dynamics}. We choose a numerical cut-off on wavenumbers ($q>10^{-8}$) to avoid any divergence at the $q=0$ mode. Aliasing of the Fourier components is prevented by implementing the $2/3$ rule~\cite{Canuto2010SpectralMF}. Next, we compute the velocity in real space [Eq.~\eqref{eq:streamF}] from
\begin{equation}
	v_{x} = \mathbf{F}^{-1} \left[ iq_{y}\psi_{\vb*{q}} \right ] ,
	\quad 
	v_{y} = \mathbf{F}^{-1} \left[ -iq_{x}\psi_{\vb*{q}} \right ]
\end{equation}
and the density dynamics [Eq.~\eqref{eq:phi_dynamics}] follows in Fourier space as
\begin{equation}\label{eq:phi_dynamics_f}
	\dot{\phi}_{\vb*{q}} = - \mathbf{F} \left[ \vb*{v} \cdot \nabla \phi\right] - a M \vb*{q}^{2} \phi_{\vb*{q}} + b \mathbf{F} \left[ \nabla^{2} \phi^{3} \right] - \kappa \vb*{q}^{4} \phi_{\vb*{q}} - i \sqrt{2MT}\vb*{q} \cdot \Lambda_{\phi \vb*{q}} .
\end{equation}
In each simulation, the system is allowed to relax and reach a steady state before taking any measurements. To probe finite-size effects, we evaluate the IEPR [Eq.~\eqref{loc_epr}] for various values of the spacial discretization and the size of the system, showing that our measurements scale like $N^2$, where $N$ is the number of lattice points, as expected [Fig.~\ref{fig:epr_dx}].

\section{Nematic director and mean curvature}\label{ap:nematic_director}

In this Appendix, we summarize the relation between symmetric traceless tensors in 2 dimensions, its principal axes, and the nematic director~\cite{DeGennes}. Let us consider the most general traceless symmetric tensor $Q_{\alpha\beta}$ in 2 dimensions:
\begin{equation}\label{eq:symtr0}
    Q_{\alpha\beta}=\begin{pmatrix}
        Q_{11} &Q_{12} \\
        Q_{12} &-Q_{11}
    \end{pmatrix}
\end{equation}
We interpret $\mathbb{Q}$ as the order parameter of a nematic field, whose orientation is $\theta_N(r,\varphi)$ at every point in space.

\subsection{Nematic director}

The eigenvalues and eigenvectors of $\mathbb Q$ [Eq.~\eqref{eq:symtr0}] read
\begin{equation}\label{eq:vpm}
	\lambda_\pm = \pm\sqrt{Q_{11}^2+Q_{12}^2} ,
	\quad
	\mathbf{v}_\pm = \left(\frac{Q_{11}\pm\sqrt{Q_{11}^2+Q_{12}^2}}{Q_{12}},1\right) .
\end{equation}
The eigenspace of $Q_{\alpha\beta}$ is defined up to a multiplication of $\mathbf{v}_\pm$ by an arbitrary scalar $s$. We are interested in the direction of the eigenvectors, so we construct the projective space associated with the eigenspace of $Q_{\alpha\beta}$ by identifying $\mathbf{v}_\pm \sim s \mathbf{v}_\pm$ for any real $s$. The eigenvectors are orthogonal, so it is sufficient to consider the direction of one of them, say $\mathbf{v}_+$ (the direction of $\mathbf{v}_-$ is orthogonal to it). In polar form, we get $\mathbf{v}_+=|\mathbf{v}_+|(\cos\theta_N,\sin\theta_N)$, where the angle $\theta_N$ can be deduced from Eq.~\eqref{eq:vpm}:
\begin{equation}
\label{eq:tan1}
    \tan\theta_N = \frac{(v_{+})_y}{(v_{+})_x}= \frac{Q_{12}}{Q_{11}+\sqrt{Q_{11}^2+Q_{12}^2}}
\end{equation}
Equation~\eqref{eq:tan1} can be used to compute $\theta_N$ if the components $Q_{\alpha\beta}$ are known. Using the trigonometric identity
\begin{equation}\label{eq:tan2}
	\tan 2 \theta_N = \frac{2 \tan \theta_N}{1-\tan^2\theta_N} ,
\end{equation}
we find
\begin{equation}\label{eq:omega1}
	\theta_N= \frac{1}{2}\tan^{-1}\frac{Q_{12}}{Q_{11}} ,
\end{equation}
yielding
\begin{equation}
    Q_{\alpha\beta}=Q_{11}\begin{pmatrix}

        1 &\tan 2 \theta_N \\
        \tan 2 \theta_N &-1
    \end{pmatrix}=\frac{Q_{11}}{\cos2\theta_N}\begin{pmatrix}
        \cos2\theta_N &\sin 2 \theta_N \\
        \sin 2 \theta_N &-\cos2\theta_N
    \end{pmatrix}
\end{equation}
Note that the relation $Q_{12}=Q_{11}\tan2\theta_N$ implies $Q_{11}^2+Q_{12}^2=Q_{11}(1+\tan^22\theta_N)=(Q_{11}/\cos2\theta_N)^2$. The final expression of the nematic tensor is therefore
\begin{equation}
    Q_{\alpha\beta}=\sqrt{Q_{11}^2+Q_{12}^2}\begin{pmatrix}
        \cos2\theta_N &\sin 2 \theta_N \\
        \sin 2 \theta_N &-\cos2\theta_N
    \end{pmatrix} ,
\end{equation}
so $S\equiv\sqrt{Q_{11}^2+Q_{12}^2}$ is the amplitude of the order parameter. In short, given a symmetric traceless tensor [Eq.~\eqref{eq:symtr0}], the directions of its principal axes are characterized by the angle $\theta_N(x,y)$ [Eq.~\eqref{eq:omega1}] under the identification $\theta_N\sim\theta_N+\pi$.

\subsection{Mean curvature}

Consider a surface $\phi$ embedded in $\mathbb{R}^3$. We use the Monge parametrization to represent a point $\mathbf{X}\in\mathbb{R}^3$ on the surface \cite{doCarmo1994}:
\begin{equation}
    \mathbf{X}(x,y)=(x,y,\phi(x,y)) .
\end{equation}
The tangent vectors $\mathbf{t}_\alpha$ ($\alpha=x,y$) and the unit normal to the surface $\mathbf{N}$ are
\begin{equation}\label{eq:N}
	\mathbf{t}_x =(1,0,\partial_x\phi) ,
	\quad
	\mathbf{t}_y = (0,1,\partial_y\phi) ,
	\quad
	\mathbf{N} = \frac{(-\partial_x\phi,-\partial_y\phi,1)}{\sqrt{1+|\nabla\phi|^2}} .
\end{equation}
The variation of $\mathbf{N}$ between two infinitesimally close points in the surface is parallel to the surface~\cite{doCarmo1994}, which leads to define the {\em extrinsic curvature tensor} $K_{\alpha}^{\beta}$ as
\begin{equation}
	\partial_\alpha \mathbf{N} = - K_{\alpha}^{\beta}\mathbf{t}_\beta ,
\end{equation}
where we use summation over repeated indices. The {\em mean curvature} follows as
\begin{equation}
	H = \frac{1}{2}K_\alpha^\alpha .
\end{equation}
By taking derivatives of $\mathbf{N}$ and projecting onto $\mathbf{t}_\beta$, we construct the tensor $K_{\alpha}^{\beta}$, and deduce
\begin{equation}\label{eq:H}
	H = \frac{(1+(\partial_y\phi)^2)(\partial_x\phi)^2-2(\partial_x\phi)(\partial_y\phi)(\partial_x\partial_y\phi) + (1+(\partial_x\phi)^2)(\partial_y\phi)^2}{2(1+|\nabla\phi|^2)^{3/2}} ,
\end{equation}
where we have used the definition of ${\bf N}$ [Eq.~\eqref{eq:N}].


\section{Derivation of the informatic entropy production rate}\label{ap:derivation_epr}

To derive the informatic entropy production rate (IEPR) [Eq.~\eqref{eq:def_epr}], we start by writing the path probability for the forward $\mathcal{P} \left(\{\vb*{J},\psi\}_0^{t} \right)$ and backward $\mathcal{P}^{R} (\{\vb*{J},\psi\}_0^{t})$ paths:
\begin{equation}
	\mathcal{P} (\{\vb*{J},\psi\}_0 ^{\tau}) = \exp(-\mathcal{A}) ,
	\quad
	\mathcal{P}^{R} \left (\{\vb*{J},\psi\}_0 ^{\tau} \right ) = \exp(-\mathcal{A}^{R}) , 
\end{equation}
where the $(\mathcal{A},\mathcal{A}^R)$ are the corresponding dynamical actions, and the time-reversal operation is defined by 
\begin{equation}
	t \xrightarrow{} \tau-t ,
	\quad
	\phi^R(\vb*{r},t)=\phi(\vb*{r},\tau-t) ,
	\quad
	\psi^R(\vb*{r},t)=-\psi(\vb*{r},\tau-t) .
\end{equation}
Given that the noise realizations in the dynamics of the density field $\phi$ [Eq.~\eqref{eq:phi_dynamics}] and the stream field $\psi$ [Eq.~\eqref{eq:stream_dynamics}] are independent, we separate $(\mathcal{A},\mathcal{A}^R)$ as
\begin{equation}
	\mathcal{A} =  \int_0^\tau \mathrm{d}t \int \, \mathrm{d}\vb*{r} \left (\mathcal{A}_\phi +\mathcal{A}_\psi \right ) ,
	\quad
	\mathcal{A}^{R} = \int_0^\tau \mathrm{d}t \int \, \mathrm{d}\vb*{r} \left(\mathcal{A}^{R}_\phi +\mathcal{A}^{R}_\psi \right) ,
\end{equation}
where $(\mathcal{A}_\phi,\mathcal{A}_\phi^R)$ and $(\mathcal{A}_\psi,\mathcal{A}_\psi^R)$ are the contributions from the dynamics $\phi$ and $\psi$, respectively. The IEPR follows as
\begin{equation}\label{eq:epr_gen}
\begin{aligned}
	\dot{\mathcal{S}} &= \lim _{\tau  \to \infty} \frac{1}{\tau } \left ( \mathcal{A}^{R}-\mathcal{A} \right )
	\\
	&	=	\underbrace{ \lim_{\tau  \to \infty} \frac{1}{\tau } \int_0^\tau \mathrm{d}t \int \mathrm{d}\vb*{r}	\left (\mathcal{A}^{R}_\phi-\mathcal{A}_\phi \right ) }_{\dot{\mathcal{S}}_{\phi}} + \underbrace{	\lim_{\tau  \to \infty} \frac{1}{\tau } \int_0^\tau \mathrm{d}t \int \, \mathrm{d}\vb*{r} \left (\mathcal{A}^{R}_\psi-\mathcal{A}_\psi \right ) }_{\dot{\mathcal{S}}_{\psi}} .
\end{aligned}
\end{equation}
The expression for $\mathcal{A}_\psi$ reads~\cite{nardini_2017}
\begin{equation}\label{eq:L}
	\mathcal{A}_\psi = \frac{1}{4\eta T}\left [\nabla^{-2} \left (L-\eta \nabla ^{4} \psi \right ) \right ]^{2} ,
	\quad
	L = (\kappa+\zeta)[(\partial_x \phi)\nabla ^2 (\partial_y \phi)-(\partial_y \phi)\nabla ^2 (\partial_x \phi)] ,
\end{equation}
and $\mathcal{A}_\phi$ is given by
\begin{equation}\label{eq:a_phi}
	\mathcal{A}_\phi = \frac{1}{4MT}\left[\nabla ^{-1} \left(\dot{\phi}+\vb*{v} \cdot \nabla \phi +\nabla \cdot \vb*{J}_{d} \right) \right]^{2} ,
	\quad
	\vb*{J}_d = - M \nabla \frac{\delta\cal F}{\delta\phi} ,	
\end{equation}
from which we deduce
\begin{equation}
	\mathcal{A}^{R}_\psi = \frac{1}{4\eta T} \left [\nabla^{-2} \left (L + \eta \nabla^{4} \psi \right) \right ]^{2} ,
	\quad
	\mathcal{A}_\phi^R = \frac{1}{4MT}\left[\nabla ^{-1} \left(\dot{\phi}+\vb*{v} \cdot \nabla \phi - \nabla \cdot \vb*{J}_{d} \right)\right]^{2} ,
\end{equation}
yielding
\begin{equation}\label{eq:s_decomp}
	\dot{\cal S}_\psi = {+} \frac{1}{T} \int \langle L \psi \rangle \, \mathrm{d}\vb*{r} ,
	\quad 
	\dot {\cal S}_\phi =-\frac 1 T \int \left\langle \left(\dot{\phi}+ \vb*{v} \cdot \nabla \phi \right) \frac{\delta\cal F}{\delta\phi} \right\rangle \, \mathrm{d}\vb*{r} ,
\end{equation}
where we have used integration by parts. In the expression of $\dot{\cal S}_\phi$, we identify a vanishing boundary term:
\begin{equation}  
	\int \, \mathrm{d}\vb*{r} \left\langle \dot \phi \frac{\delta\cal F}{\delta\phi} \right\rangle = \left\langle \frac{\mathrm{d}\cal F}{\mathrm{d} t} \right\rangle = \underset{\tau\to\infty}{\lim} \frac{{\cal F}(\tau) - {\cal F}(0)}{\tau} = 0 .
\end{equation}
Using the definition $v_\alpha = \epsilon_{\alpha\beta}\partial_\beta\psi$ [Eq.~\eqref{eq:streamF}] and the expression of the free energy $\cal F$ [Eq.~\eqref{eq:free}], the remaining contribution to $\dot{\cal S}_\phi$ can be written as
\begin{equation}
\begin{aligned}
	T \dot{\cal S}_\phi &= {-}\epsilon_{\alpha\beta} \int \langle (\partial_\beta \psi) (\partial_\alpha \phi) ( dg/d\phi - \kappa\nabla^2 \phi ) \rangle \,\mathrm{d}\vb*{r}
	\\
	&={-} \epsilon_{\alpha\beta} \int \langle (\partial_\beta \psi) (\partial_\alpha g - \kappa (\partial_\alpha \phi) \nabla^2 \phi ) \rangle \,\mathrm{d}\vb*{r}
	\\
	&= {+} \epsilon_{\alpha\beta} \int \langle \psi (\partial_\beta\partial_\alpha g - \kappa \partial_\beta( (\partial_\alpha \phi) \nabla^2 \phi ) ) \rangle \,\mathrm{d}\vb*{r} ,
\end{aligned}
\end{equation}
where we have integrated by parts. Given that {$\epsilon_{\alpha\beta} \partial_\alpha\partial_\beta h(\phi) =0$} for an arbitrary function $h$ (since $\epsilon_{\alpha\beta}=-\epsilon_{\alpha\beta}$), we deduce
\begin{equation}\label{eq:s_phi}
	T \dot{\cal S}_\phi = {-} \kappa \epsilon_{\alpha\beta} \int \langle \psi (\partial_\alpha \phi) \nabla^2 (\partial_\beta \phi) \rangle \,\mathrm{d}\vb*{r} .
\end{equation}
Moreover, using the definition of $L=(\kappa+\zeta)\epsilon_{\alpha\beta}(\partial_\alpha\phi) \nabla^2(\partial_\beta \phi)$ [Eq.~\eqref{eq:L}] and $\dot{\cal S}_\psi $ [Eq.~\eqref{eq:s_decomp}], we find
\begin{equation}\label{eq:s_psi}
	T \dot{\cal S}_\psi = {+}(\zeta + \kappa) \epsilon_{\alpha\beta} \int 
    \langle  \psi (\partial_\alpha\phi) \nabla^2(\partial_\beta \phi)\rangle \, \mathrm{d}\vb*{r} .
\end{equation}
Comparing Eqs.~\eqref{eq:s_phi} and~\eqref{eq:s_psi}, we find $\dot{\cal S}_\phi = -\kappa(\zeta + \kappa)^{-1} \dot{\cal S}_\psi$, from which we deduce the total IEPR:
\begin{equation}
	T\dot{\cal S} = T (\dot{\cal S}_\phi + \dot{\cal S}_\psi) = \frac{\zeta}{\zeta+\kappa} \int \langle L \psi \rangle \,\mathrm{d} \vb*{r} .
\end{equation}
In the noiseless regime ($T\to0$), the dynamics of $\psi$  [Eq.~\eqref{eq:stream_dynamics}] reduces to $L=\eta \nabla^{4}\psi$, yielding
\begin{equation}
\label{eq:EPRderivation}
	T\dot{\cal S} = \frac{\eta \zeta}{\zeta+\kappa} \int \langle (\nabla^2\psi)^2 \rangle \, \mathrm{d} \vb*{r} ,
\end{equation}
which coincides with Eq.~\eqref{loc_epr}.


\section{Derivation of the dissipation rate}\label{app:LIT}

To derive the relation between the dissipation rate $T\dot{\cal S}_{\rm tot}$ [Eq.~\eqref{eq:diss}] and IEPR $\dot{\cal S}$ [Eq.~\eqref{loc_epr}], we start by expressing $T\dot{\cal S}_{\rm tot}$ as the sum of products between fluxes and forces~\cite{Mazur}:
\begin{equation}\label{eq:EPRfluxforce}
	T\dot{\cal S}_\mathrm{tot} = \int ({\mathbb J}\cdot{\mathbb X}) \mathrm{d} \vb*{r} = \int \left[ -J_\alpha \partial_\alpha \frac{\delta\cal F}{\delta\phi} + v_{\alpha\beta} \left( \Sigma^\mathrm{tot}_{\alpha\beta}-\Sigma^\mathrm{P,tot}_{\alpha\beta}\right) + r \Delta\mu \right]  \mathrm{d} \vb*{r} ,
\end{equation}
where we have used the definitions of $({\mathbb X}, {\mathbb J})$ [Eq.~\eqref{eq:lit}]. Integrating by parts the first and second terms in Eq.~\eqref{eq:EPRfluxforce}, we deduce
\begin{equation}
\begin{aligned}
	T\dot{\cal S}_\mathrm{tot} &= \int \left[ (\partial_\alpha J_\alpha) \frac{\delta\cal F}{\delta\phi} + v_{\beta} \left( \partial_\alpha\Sigma^\mathrm{P,tot}_{\alpha\beta} - \partial_\alpha\Sigma^\mathrm{tot}_{\alpha\beta} \right) + r \Delta\mu  \right]  \mathrm{d} \vb*{r}
	\\
	&= \int \left[ - \frac{\delta\cal F}{\delta\phi} (\partial_t\phi + v_\alpha\partial_\alpha\phi) + v_{\beta}\partial_\alpha\Sigma^\mathrm{P,tot}_{\alpha\beta} + r \Delta\mu \right] \mathrm{d} \vb*{r} ,
\end{aligned}
\end{equation}
where we have used Eqs.~\eqref{eq:phi_dynamics} and~\eqref{eq:stokes_bis}. Using the following identity for the passive stress~\cite{Cates2025}
\begin{equation}
	\phi\,\partial_\alpha\frac{\delta\cal F}{\delta\phi} = - \partial_\alpha\Sigma^\mathrm{P,tot}_{\alpha\beta} ,
\end{equation}
together with
\begin{equation}
	\int \frac{\delta \mathcal{F}}{\delta\phi} \, \partial_t\phi\, \mathrm{d} \vb*{r} = - \frac{\mathrm{d}\mathcal{F}}{\mathrm{d}t} ,
	\quad
	\partial_\alpha v_\alpha = 0 ,
\end{equation}
we then deduce
\begin{equation}\label{eq:EPRsplit}
	T\dot{\cal S}_\mathrm{tot} = - \frac{\mathrm{d}\mathcal{F}}{\mathrm{d}t}  + \int  ( r \Delta\mu ) \mathrm{d} \vb*{r} .
\end{equation}
As mentioned in the main text, the term $\mathrm{d}\mathcal{F}/\mathrm{d}t$ vanishes in steady state. Combining the definitions of the chemical rate $r$ [Eq.~\eqref{eq:r}] and of the determinant of the Onsager matrix $\det\mathbb L$ [Eq.~\eqref{eq:det}], we obtain
\begin{equation}\label{eq:rDmu}
	\int (r\Delta\mu) {\rm d}\vb*{r} = \frac{(2\eta)^4\Delta\mu ^2}{M}\int ( \det\mathbb{L} )\mathrm{d} \vb*{r} - \int v_{\alpha\beta} \Sigma_{\alpha\beta}^A {\rm d}\vb*{r} .
\end{equation}
Using the definition $v_\alpha = \epsilon_{\alpha\beta}\partial_\beta\psi$ [Eq.~\eqref{eq:streamF}] and integrating by parts, we write the last term in Eq.~\eqref{eq:rDmu} as follows:
\begin{equation}\label{eq:vS}
	\int v_{\alpha\beta}\Sigma_{\alpha\beta}^A \mathrm{d}\vb*{r} = \int \epsilon_{\beta\gamma}\partial_\alpha\partial_\gamma \psi \Sigma_{\alpha\beta}^A \mathrm{d}\vb*{r} = \int \psi\,\epsilon_{\beta\gamma}\partial_\alpha\partial_\gamma  \Sigma_{\alpha\beta}^A \mathrm{d}\vb*{r} .
\end{equation}
Finally, using $\Sigma_{\alpha\beta}=\Sigma_{\alpha\beta}^P+\Sigma_{\alpha\beta}^A$ and Eq.~\eqref{eq:stress_tensor}, we can rewrite the dynamics of $\psi$ [Eq.~\eqref{eq:stream_dynamics}] in terms of the active stress:
\begin{equation}\label{eq:stokes2}
	\eta \nabla^4 \psi = - \epsilon_{\beta\gamma} \frac{\kappa+\zeta}{\zeta}\partial_\alpha\partial_\gamma \Sigma_{\alpha\beta}^A .
\end{equation}
Substituting Eq.~\eqref{eq:stokes2} into Eq.~\eqref{eq:vS}, we find that Eq.~\eqref{eq:vS} is identical to the IEPR of AMH:
\begin{equation}\label{eq:connectionIEPR}
	\int v_{\alpha\beta}\Sigma_{\alpha\beta}^A \mathrm{d}\vb*{r} = \frac{\eta\zeta}{\kappa+\zeta} \int (\psi\, \nabla^4\psi) \mathrm{d}\vb*{r} = \frac{\eta\zeta}{\kappa+\zeta} \int (\nabla^2 \psi)^2  \mathrm{d}\vb*{r} = T\dot{\cal S} \, ,
\end{equation}
where we have identified the expression of IEPR $\dot{\cal S}$ [Eq.~\eqref{loc_epr}]. Finally, substituting the relations from Eqs.~\eqref{eq:rDmu} and~\eqref{eq:connectionIEPR} into Eq.~\eqref{eq:EPRsplit}, we obtain the connection between the steady-state dissipation $T\dot{\cal S}_{\rm tot}^{\,(\mathrm{ss})}$ and $\dot{\cal S}$ in Eq.~\eqref{eq:EPRsplit2Main}. Last, we note that combining the definition of $\det\mathbb L$ [Eq.~\eqref{eq:det}] with the following relation
\begin{equation}
	\tilde{\Sigma}^A_{\alpha\beta}\tilde{\Sigma}^A_{\alpha\beta} = \frac{\zeta^2}{2\Delta\mu^2}|\nabla\phi|^4 ,
\end{equation}
we can simplify Eq.~\eqref{eq:EPRsplit2Main} as
\begin{equation}\label{eq:resEPR}
	T ( \dot{\cal S}_\mathrm{tot}^{\,(\mathrm{ss})} - \dot{\cal S} ) = \int  \left(c_r\Delta\mu ^2 - \frac{\zeta^2}{2\eta} |\nabla\phi|^4\right) \mathrm{d} \vb*{r} .
\end{equation}
Therefore, any local inhomogeneity of the swimmer density $\phi$ tends to reduce $\dot{\cal S}_\mathrm{tot}^{\,(\mathrm{ss})} - \dot{\cal S}$.


\section{Independence of IEPR from boundary conditions}\label{app:bc}

\begin{figure}
	\centering
	\includegraphics[width=\linewidth]{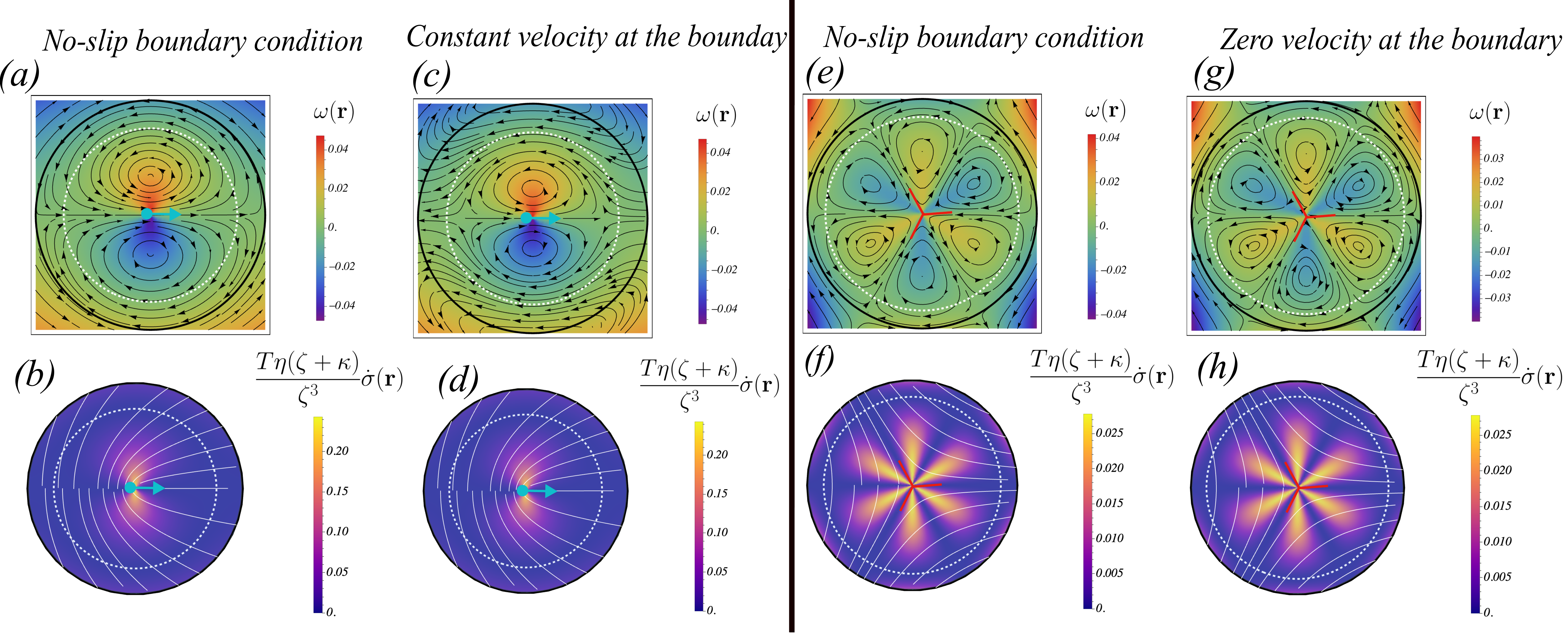}
	\caption{Analytical prediction for the local IEPR and velocity fields around isolated defects topological: for a $+\frac 1 2$ defect, (a-b)~no-slip boundary and (c-d)~constant-velocity boundary; for a $-\frac 1 2$ (e-f)~no-slip boundary and (g-h)~constant-velocity boundary. Parameters: $\kappa=0.1$, $\zeta=0.16$, $\eta=1.67$, and $\xi\gg R$ (no-slip limit). The white dashed circles indicate the inner region $r<r_C$ wherein results are  expected to be universal and insensitive to the boundary conditions.
	}
	\label{fig:epr_bc}
\end{figure}

In this Appendix, we demonstrate that close to the defect the profile of IEPR [Eq.~\eqref{eq:sig}] does not depend on the specific boundary condition. In fact, this result stems from the fact that the effects of boundary conditions decay within a boundary layer of thickness $\delta$. We start from the stream function of a $+\frac 1 2$ defect: $\psi=\bar{\psi}_{+\frac{1}{2}}+\psi_0$ [Sec.~\ref{sec:isolatedDefects}], with $\bar{\psi}_{+\frac{1}{2}}$ given by Eq.~\eqref{eq:sfPlus} and $\psi_0 = (a_1r+b_1r^3)\sin\varphi$. We impose a constant velocity at $r=R$ directed along the $x-$axis:
\begin{equation}\label{eq:G1}
	v_x(R,\varphi) = v_R,
	\quad
	v_y(R,\varphi)=-\partial_x\psi(R,\varphi) = 0 ,
\end{equation}
where $v_x(r,\varphi)=\partial_y\psi(r,\varphi)$ and $v_y(r,\varphi)=-\partial_x\psi(r,\varphi)$ can be calculated easily~\cite{Giomi2014}:
\begin{align}\label{eq:G2}
	&v_x(r,\varphi)=a_1+b_1r^2(2-\cos2\varphi)+\frac{\zeta}{12\eta}\left[ 3(R-r)+r\cos2\varphi\right] ,
	\\
	&v_y(r,\varphi)=-b_1r^2\sin2\varphi + \frac{\zeta}{12\eta} r \sin2\varphi .
\end{align}
Substituting Eq.~\eqref{eq:G2} into Eq.~\eqref{eq:G1}, we determine the two coefficients $(a_1,b_1)$:
\begin{equation}
	a_1=v_R-\frac{\zeta R}{6 \eta},
	\quad
	b_1=\frac{\zeta}{12R\eta}.
\end{equation}
Correspondingly, the stream function satisfying the boundary conditions is
\begin{equation}\label{eq:G3}
	\psi_{+\frac{1}{2}} = \left(v_R + \frac{\zeta}{12R\eta}(r-R)^2 \right) r \sin\varphi .
\end{equation}
Substituting Eq.~\eqref{eq:G3} into Eq.~\eqref{eq:loc_epr2}, we find that the IEPR is independent of $v_R$, since $\nabla^2(r\sin\varphi)=0$:
\begin{equation}\label{eq:G4}
	 T\dot{\sigma}_{+\frac{1}{2}}  (\vb*{r}) = \frac{\zeta^3}{\eta(\kappa+\zeta)}\biggl[\frac{1}{2}-\frac{2}{3}\frac{r}{R} \biggl]^2\sin^2\varphi .
\end{equation}
We note that the differences between Eq.~\eqref{eq:G4} and  Eq.~\eqref{eq:sig} should be localized in a boundary layer of thickness $\delta$. We expect that, in the bulk ($r<r_C\simeq R-\delta$), the solution is only mildly dependent on the specific boundary conditions. We estimate $r_C$ as the radius where the IEPR is minimum. Imposing $\partial_r \dot{\sigma}_{+\frac{1}{2}} (r_C,\varphi)=0$ on Eq.~\eqref{eq:G4}, we find $r_C=3R/4$. Figures~\ref{fig:epr_bc}(a,c) [resp. (b,d)] compare the vorticity [resp. the IEPR] of an isolated $+\frac 1 2$ defect. The dashed white circle indicates $r=r_C$. While the solutions are different at $r=R$ (black circle), they are almost identical in the bulk ($r<r_C$) thus supporting the claim that IEPR has a universal profile close to the defects' center.

Following the same procedure and imposing $v_R=0$ [Eq.~\eqref{eq:G1}] on the velocity of a $-\frac 1 2$ defect, we find the following expressions for the constants:
\begin{equation}
	a_1 = \frac{7R\alpha }{240\eta R^2} ,
	\quad
	b_1 = - \frac{R\alpha }{60 \eta R^4} .
\end{equation}
The corresponding expression of the IEPR is
\begin{equation}
\label{eq:eprBCminus}
    T\dot{\sigma}_{-\frac{1}{2}}  (\vb*{r}) = \frac{\zeta^3}{\eta(\kappa+\zeta)}\biggl[\frac{1}{6}-\frac{5}{15}\left(\frac{r}{R}\right)^3 \biggl]^2\sin^2(3\varphi) .
\end{equation}
The comparison between Eq.~\eqref{eq:eprMinusHalf}  (no-slip BCs) and  Eq.~\eqref{eq:eprBCminus} is shown in Fig.~\ref{fig:epr_bc}(f) and (h), respectively. The panels Fig.~\ref{fig:epr_bc}(e) and (g) show the corresponding vorticity and flow fields. Similarly to the case of the $+\frac 1 2$ defect, the solutions are almost identical within a radius $r_C=5^{1/3}R/2$, as determined by imposing $\partial_r \dot{\sigma}_{-\frac{1}{2}}(r_C,\varphi)=0$.

\section{Spatial distribution of the active force}\label{app:active_force}
\begin{figure}
    \centering
    \includegraphics[width=\linewidth]{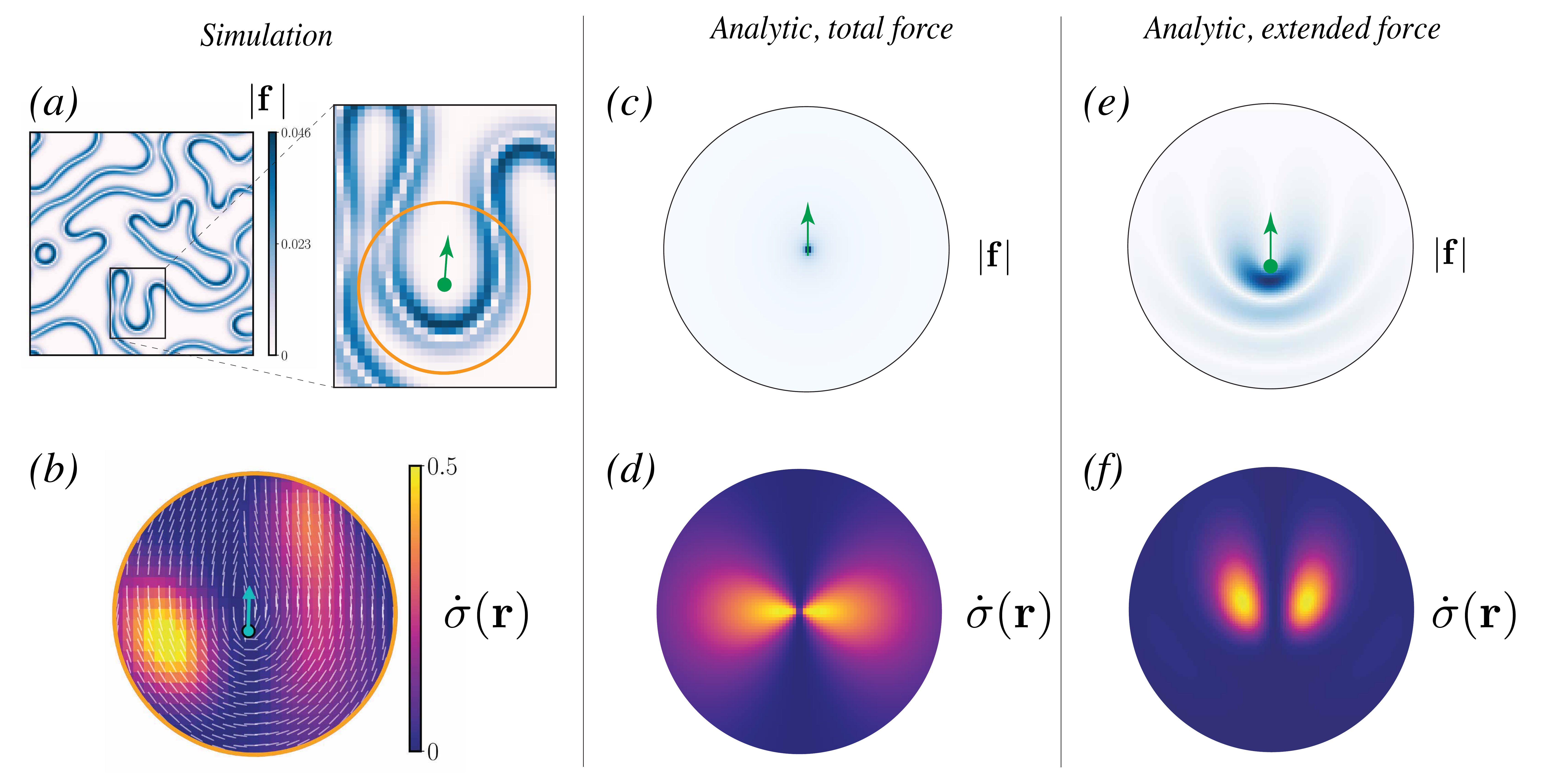}
    \caption{{Prediction of the qualitative localization of the IEPR derived from various force fields. The plots are in arbitrary units. (a,b) Exact force field and corresponding IEPR form numerical simulations. Panel (c): Analytic force field around an isolated $+\frac{1}{2}$ defect (analog to active nematics, \cite{Giomi2014}). Note that the force is maximal at the center of the defect, contrary to (a). Consequently, while the mirror symmetry is qualitatively preserved, the IEPR [Panel (d)] is maximal at the center, and there is no separation between the two lobes of IEPR. Panel (e): Analytic ansatz for the force field around a $+\frac{1}{2}$ defect, which qualitatively reproduces the symmetries of [Panel (a), inset] as well as the localization along the interface. The corresponding IEPR [Panel (f)] has two localized lobes consistent with the full solution shown in Panel (b).}} 
    \label{fig:appG_ActForce}
\end{figure}

{
In this Appendix, we comment on the relation between the local IEPR $\dot\sigma$ [Eq.~\eqref{eq:loc_epr2}] and the distribution of the total force $f_\alpha \propto \partial_\beta Q_{\alpha\beta}$ [Eq.~\eqref{eq:fa}] around \textit{individual} defects by comparing numerical simulations with some analytical profiles. In the turbulent state, we consider a specific region with a $+\frac 1 2$ defect, whose orientation is indicated with a green arrow [Fig.~\ref{fig:appG_ActForce}(a)]. The corresponding IEPR is maximal on the two sides of the defect [Fig.~\ref{fig:appG_ActForce}(b)], is approximately symmetric with respect to reflections across the defect's orientation, and is comparatively lower near the origin. In what follows, we provide some arguments to reproduce the profiles of $\dot\sigma$ and $f_\alpha$ around the defect.
}



{
In Sec.~\ref{sec:isolatedDefects}, our analytic profile of the local IEPR $\dot\sigma$ around a point-like defect [Eq.~\eqref{eq:eprPlusHalf}] assumes that the defect is isolated and the core size is zero, namely the amplitude of the $Q-$tensor is constant everywhere except at the origin [\ref{ap:nematic_director}]. As a result, the corresponding force profile $|{\bf f}|$ is strongly localized at the defect core [Fig.~\ref{fig:appG_ActForce}(c)], and $\dot\sigma$ is maximal at the defect core with a mirror symmetry with respect to the defect orientation [Fig.~\ref{fig:appG_ActForce}(d)], in contrast with the numerical profiles [Figs.~\ref{fig:appG_ActForce}(a-b)].
}


{
We now show that the lobed-structure of Figs.~\ref{fig:appG_ActForce}(a-b) is related to the non-constant amplitude of the $Q-$tensor close to the defect core. We propose an ansatz for the force field $\hat{\mathbf{f}}(\mathbf{r})$ that qualitatively mimics the geometry of the numerical profile:}
\begin{equation}\label{eq:auxF}
    \hat{\mathbf{f}}(\mathbf{r}) = e^{-r}\nabla \chi(\vb{r}) ,
    \quad
    \chi=\frac{1}{2}\left[1-\cos \biggl(2 r \cos \frac{\theta}{2}  \biggl)\right] ,
\end{equation}
{where we assume that the defect is at the origin of the coordinate system, $r=\sqrt{x^2+y^2}$, and $\theta = \arctan (y/x)$. The ansatz in Eq.~\eqref{eq:auxF} is localized along the interface, vanishes at the defect core, and has a mirror symmetry [Fig.~\ref{fig:appG_ActForce}(e)], as seen in the numerical solution [Fig.~\ref{fig:appG_ActForce}(a)]. The corresponding IEPR profile follow by using Eqs.~\eqref{eq:loc_epr2} and~\eqref{eq:134}: it features two well-separated lobes located symmetrically with respect to the defect orientation [Fig.~\ref{fig:appG_ActForce}(f)] in agreement with simulations [Fig.~\ref{fig:appG_ActForce}(b)].}

{
In short, we have shown that, for an individual defect, the geometry of the IEPR profile $\dot\sigma$ is encoded in the details of the corresponding force profile $|{\bf f}|$ close to the defect core. For defect pairs [Sec.~\ref{sec:pair}], the deviation between our analytics (that rely on point-defect assumption) and numerics is less severe [Figs.~\ref{fig:epr}(d-g)]. In this case, the constructive interference of vorticities occurs along the line connecting the two defects centers: the profile $\dot\sigma$ is overwhelmingly dominated by these vorices, so the details close to defect core are largely irrevelant. 
}


\section*{References}

\bibliographystyle{iopart-num}
\bibliography{references}

\end{document}